\documentclass[11pt,notoc]{JHEP3}
\usepackage{epsfig}
\usepackage{amsmath}

\usepackage{setspace}

\newcommand{\be}{\begin{equation}}
\newcommand{\ee}{\end{equation}}
\newcommand{\bea}{\begin{eqnarray}}
\newcommand{\eea}{\end{eqnarray}}

\newcommand{\non}{\nonumber \\}

\def\half{{1 \over 2}}

\def\pa{{\partial}}

\def\({\left(} \def\){\right)}
\def\[{\left[} \def\]{\right]}

\def\cC{{\cal C}}
\def\O{{\cal O}}
\def\L{{\cal L}}
\def\R{{\cal R}}
\def\J{{\cal J}}

\def\gtt{g_{tt}}
\def\gtr{g_{tr}}
\def\grr{g_{rr}}

\def\al{\alpha}
\def\bt{\beta}
\def\gm{\gamma}
\def\lam{\lambda}

\def\eps{\epsilon}
\def\th{\theta}

\preprint{   AEI-2009-077}

\title{\center{ Generalized harmonic formulation in spherical symmetry }}

\author{Evgeny Sorkin \\
Max-Planck Institute for Gravitational Physics, Albert Einstein Institute,\\
Am Muehlenberg 1, 14476, Golm, Germany \\
E-mail: \email{evgeny@aei.mpg.de} }

\author{Matthew W. Choptuik \\
Department of Physics and Astronomy, University of British Columbia \\
6224 Agricultural Road, Vancouver BC, Canada, V6T 1Z1 \\
CIFAR Cosmology and Gravity Program \\
E-mail: \email{choptuik@physics.ubc.ca}}

\abstract{In this pedagogically structured article, we describe 
a generalized harmonic formulation of the Einstein equations 
in spherical symmetry which is regular at the origin.
The generalized harmonic approach has attracted significant attention 
in numerical relativity over the past few years, especially as applied 
to the problem of binary inspiral and merger.
A key issue  when using the technique is the choice of the gauge
source functions, and recent work has provided several prescriptions
for gauge drivers designed to evolve these functions
in a controlled way. We numerically investigate
the parameter spaces of some of these drivers in the context
of fully non-linear collapse of a real, massless scalar field, 
and determine nearly optimal parameter settings for specific situations. 
Surprisingly, we find that many of the drivers that perform well in 3+1 
calculations that use 
Cartesian coordinates, are considerably less effective in spherical symmetry, 
where some of them are, in fact, unstable.}

\begin{document}

%%%%%%%%%%%%%%%%%%%%%%%%%%%%%%%%%%%%%%%%%%%%%%
\section{Introduction}
\label{sec_intro}
%%%%%%%%%%%%%%%%%%%%%%%%%%%%%%%%%%%%%%%%%%%%%%
Solving Einstein equations numerically is a notoriously difficult
task. After many years of research, several well-posed formulations of
the Einstein equations have been proposed and tested. These include 
constrained Arnowitt-Deser-Misner (ADM) \cite{ADM,York}, hyperbolic
Baumgarte-Shapiro-Shibata-Nakamura (BSSN) \cite{BSSN} and characteristic
evolution \cite{char_evol}, just to name a few: we refer the 
reader to \cite{Reula,Lehner,PretoriusBH} for reviews of these and other 
approaches.  
Among the ingredients 
that are key to the success of any particular formulation are 
1) an appropriate choice of dynamic
variables that results in a well-posed system, and 2) a choice of
coordinates that remain regular during the course of the evolution. 
In this paper we focus on a specific well-posed approach
known as the generalized harmonic (GH) formulation.  This form of 
the Einstein equations has recently
attracted significant attention in the numerical relativity community, in large 
part because of its use in obtaining the first long-term
evolution of binary black-hole inspiral and merger~\cite{FP1,FP2,FP3}.

In essence, the GH approach is a way to write the field equations such
that the resulting system is manifestly hyperbolic, taking the form of
a set of quasi-linear wave equations for the metric components. 
The basic idea underlying the strategy has a long and distinguished history: 
specifically, the use of harmonic coordinates has been instrumental
in establishing many 
fundamental results in General Relativity (GR) including the
characteristic structure of the theory~\cite{deDonder}, and the
well-posedeness of the Cauchy problem for Einstein's equations
\cite{bruhat,FisherMarsden}.
However, from the computational 
point of view, harmonic gauge\footnote{In this
  paper ``gauge'' means ``coordinate choice'', and we use both
  expressions interchangeably.}  can be too restrictive, and numerical
implementations using it may develop coordinate
pathologies, as described, for instance, in \cite{AlcubierreMasso} and \cite{Teukolsky}.
More recently, it was realized by Friedrich
\cite{Friedrich85}, and independently by Garfinkle
\cite{Garfinkle2001}, that much of the coordinate freedom apparently 
lost by the specific choice of harmonic gauge could be regained 
through the introduction of certain gauge source functions, while 
at the same time maintaining the desirable property of strong 
hyperbolicity of the field equations.
In fact, the source functions can be thought
of as representing the coordinate freedom of the Einstein equations,
and when constructing solutions of the equations, via an initial value
approach, for example, they must be completely specified in some fashion.

Following Garfinkle's pioneering use of the generalized harmonic approach 
in his study of generic singularity formulation in cosmologies with scalar 
field matter \cite{Garfinkle2001}, 
the technique was successfully employed by Pretorius
\cite{FP1,FP2,FP3}, and subsequently by others
\cite{Lindblometal2,Scheel_etal,LSUgroup}, for simulations of binary
black hole coalescence.  However, the total number of physical
scenarios studied so far using the GH approach is limited, and there 
is an argument to be made for a more systematic 
exploration of the method's potential.
This is especially the case given the relative lack of proven 
prescriptions for choosing the gauge functions appropriately in instances 
where the gravitational field is highly nonlinear and dynamic.
Moreover, in order to expedite experimentation with the approach, we feel 
that it is useful to start with systems with a high degree of symmetry. 
Restriction to highly symmetric spacetimes reduces the effective spatial 
dimensionality of the partial differential equations that must be solved, 
yields algebraically simpler equations, and, overall, leads to enormous 
savings in the computational resources required to simulate a single 
spacetime.  This in turn allows for much more detailed and thorough
surveys of the multi-dimensional parameter spaces that typically arise 
from a given choice of gauge functions.

In this paper, then, we focus on the application of the generalized 
harmonic approach to the problem of gravitational collapse in
spherically symmetric $D$-dimensional spacetime.  Even with the restriction to spherical symmetry,
we find that the strong-field aspects of the collapse process 
present significant challenges regarding the choice of the gauge 
functions.  Ironically, some of these challenges may in
fact be related to the symmetry restriction itself.
As usual, in situations where a black hole forms,
care must be taken to avoid the central singularity.  This can 
be done through the use of singularity-avoiding coordinates, by
excising the singularity from the computational domain, or with
a combination of both strategies.  Within the context of the GH formulation
any such strategy must also be preserve the strong hyperbolicity
of the field equations.

Although we view our study of the GH approach for spherically symmetric 
collapse as interesting in its own right, a primary goal
of this research is to prepare for
an investigation of axially symmetric systems using an analogous formulation.  
We thus consider our spherically-symmetric set up as a
valuable toy model for the phenomenologically richer axisymmetric
situation. In both cases it is natural to use coordinates in
which the symmetries of the spacetime are explicit. These coordinates,
however, are formally singular: at the origin in spherical symmetry,
and on the axis in axial symmetry.  Thus, in both instances the field equations have 
to be regularized in numerical implementations,  and one of the 
results of our work is a regularization procedure that is compatible with the GH approach. 
Moreover, we expect that the experience gained from our spherically symmetric calculations
concerning how to choose gauge source functions will also prove useful for the more 
general case of axisymmetric computations.

In order to maximize the usefulness of this paper to other researchers 
interested in experimenting with the generalized harmonic approach, we have 
attempted to make the following presentation reasonably self-contained and 
pedagogical in nature.
We thus begin in Sec.~\ref{sec_GH-formulation} with a brief presentation of the basic GH 
formulae in full generality, along with a discussion of the constraint
equations.  Although the constraints are consistently preserved by the
GH evolution equations in the continuum limit, in numerical calculations at
finite resolution, deviations from the constraints generically develop.  In order 
to maintain stability these deviations must be damped and we describe a 
method that effectively achieves this damping. Sec.
\ref{sec_gauge_conditions} is devoted to a detailed discussion of coordinate
conditions.  One key issue that we consider is the non-trivial problem of 
prescribing the GH source functions to mimic some of the more 
popular and successful coordinate conditions that have historically been
used in numerical relativity calculations.
Following recent proposals
\cite{FP2,FP3,Lindblometal2,Scheel_etal} we describe the formulation of the gauge
conditions as hyperbolic evolution equations: is this approach the gauge functions 
are evolved, or ``driven'', to desired targets in a controlled way, rather than being fixed 
instantly.

In Sec.~\ref{sec_formulation} we adapt the GH formulae to the
case of asymptotically flat, spherically symmetric  configurations in $D$ spacetime dimensions.  We derive
the field equations, cast them into a form suitable for numerical
solution, discuss initial and boundary conditions, and regularize
the singular origin by introducing a new variable.  Since we use
spherical coordinates adopted to the symmetry, the GH source functions
appear to diverge at the origin as $1/r$.  Hence, we regularize these 
functions as well by subtracting off the singular contribution that appears in the 
flat spacetime limit.  The operators that appear in the various gauge drivers 
then act on the regularized source functions.

In order to endow our model with non-trivial dynamics, we introduce
a minimally coupled, real, massless scalar field. 
The initial distribution of
the scalar matter is freely specified and our results are grouped
according to the ``strength'' of the initial data.  In each case we
simulate the time evolution of a single Gaussian pulse of scalar
field that is initially centered at the origin.  The weak and the
intermediate data correspond to the dispersion of relatively dilute
pulses, while a typical strong data configuration collapses to form a black hole or 
nearly does so.

Mathematically, the task of treating the coupled Einstein-scalar system
involves the solution of a 
set of several quasi-linear wave equations. (Here we note that some of 
the gauge drivers involve auxiliary variables that obey first-order-in-time
differential equations.)  Our numerical approach to solving this
system using finite difference techniques is detailed in 
Sec.~\ref{sec_evolution}.  We compactify the
spatial (radial) dimension into a finite region and cover it by a discrete
lattice. This allows us to include spatial infinity on the finite difference
mesh, which has the advantage of enabling us to set exact boundary 
conditions corresponding to asymptotic flatness.
Following \cite{FP1,FP2} we directly discretize the second-order-in
time-wave-equations on the mesh, and use a point-wise Gauss-Seidel
relaxation method to update the discrete unknowns at each time
step.  In order to damp high-frequency components of the numerical
solution---which can generically lead to instabilities---we incorporate 
explicit dissipation of the Kreiss-Oliger type.~\cite{KO}
This dissipation is also essential for 
attenuating spurious reflections from the outer region of the compactified
domain that would otherwise quickly contaminate the solution in the 
interior (i.e.~near the origin).

For the case of black hole formation we have investigated both of the 
approaches mentioned above for avoiding the central physical singularity.
On the one hand, we have implemented an excision technique, in which an
excision surface is chosen so that all characteristics on it are
pointing inwards, obviating the need for explicit boundary conditions
for the evolution equations. 
On the other hand, we have also experimented with the use 
of singularity avoiding slicing
conditions, that ``freeze'' the evolution in the strong curvature regions.
However, we find that in our case  the calculations using singularity-avoiding 
slicings tend not to run as long as those with excision and appear to crash prematurely due to
numerical errors that build up in the strong curvature regions.

Sec.~\ref{sec_results} is devoted to a discussion of our 
detailed investigation of the performance of several coordinate 
conditions as applied to calculations involving various strengths
of initial data.  As already mentioned, the
parameter spaces associated with many of the gauge drivers that we 
consider here are multidimensional. Thus, even with the significant 
reduction in needed computational resources that the restriction
to spherical symmetry provides, we have not found it feasible to 
identify optimal parameters in all cases.  In some instances then, we 
simply report what appears to be typical behavior for a particular gauge, 
while still trying to explore the effects of the variation of key parameters 
on the quality of the solutions.  Interestingly, we find that several of the 
gauge drivers that have been successfully used in the $3+1$ simulations 
of black hole collisions that use Cartesian 
coordinates~\cite{FP2,FP3,Lindblometal2,Scheel_etal} are considerably 
less effective for our spherically symmetric calculations.
In particular, it is not always possible to drive the
lapse to a certain value as reported in \cite{FP2,FP3}, nor is it 
always possible to enforce a desired gauge for a long time by using
one of the drivers described in \cite{Lindblometal2}. 
Overall, our calculations seem to be more sensitive to the specific 
choices of parameters for the drivers than the Cartesian computations,
and this is an issue which warrants further investigation.

Nevertheless, our results indicate that with a certain amount of parameter
tuning, several of the gauge conditions that we investigate facilitate 
the simulation of many interesting scenarios.  We are thus encouraged 
by this particular application of the generalized harmonic approach, 
and our conclusions and discussion in 
Sec.~\ref{sec_conclusions} includes an outline of some
future extensions of the work.

%%%%%%%%%%%%%%%%%%%%%%%%%%%%%%%%%%%%%%%%%%%%%%%%%%%%%%%%%%%%
\section{Generalized harmonic formulation}
\label{sec_GH-formulation}
%%%%%%%%%%%%%%%%%%%%%%%%%%%%%%%%%%%%%%%%%%%%%%%%%%%%%%%%%%%%
We consider the Einstein equations on a $D$-dimensional spacetime and
written in the form
\be
\label{Eeqs}
R_{\mu\nu}=8 \pi G_N \bar{T}_{\mu\nu}\equiv 8 \pi G_N
\(T_{\mu\nu}-\frac{1}{D-2} g_{\mu\nu} T \), \ee
where $g_{\mu\nu}$ is the metric, $R_{\mu\nu}$ is the Ricci tensor,
$T_{\mu\nu}$ is the energy-momentum tensor of the matter with trace
$T$, and $G_N$ is the $D$-dimensional Newton constant.  Hereafter, we
adopt units for which $8 \pi G_N=1$.

The Ricci tensor that appears in the left-hand-side of (\ref{Eeqs})
contains various second derivatives of the metric components
$g_{\mu\nu}$: these second derivatives collectively constitute the
principal part of $R_{\mu\nu}$, viewed as an operator on $g_{\mu\nu}$.
This principal part can be decomposed into a term $g^{\al\bt}
\pa_{\al\bt} g_{\mu\nu}$, plus mixed derivatives of the form $g^{\al
  \gm} \pa_{\al\mu} g_{\gm\nu}$. Without the mixed derivatives,
(\ref{Eeqs}) would represent manifestly (and strongly) hyperbolic wave
equations for the $g_{\mu\nu}$~\cite{Friedrich96}.  Strong
hyperbolicity is a highly desirable property since mathematical
theorems then ensure (local) existence and uniqueness of solutions at
the continuum level.  This, in turn, means that it should be possible
to construct stable (convergent) numerical discretizations of the
field equations.

One can view the generalized harmonic (GH) formulation of general
relativity as a particular method that eliminates the mixed second
derivatives appearing
in~(\ref{Eeqs})~\cite{Friedrich85,Garfinkle2001,FP1,FP3,Lindblometal1}.
As the name suggests, the technique generalizes the harmonic approach
in which the spacetime coordinates, $x^\mu$, satisfy the harmonic
coordinate condition
\be
\label{Harmonic}
\Box x^\al =0. \ee
Here we have
\be
\label{box_x}
\Box x^\al=\frac{1}{\sqrt{-g}} \pa_\nu \(\sqrt{-g} g^{\al\nu}\)
=-\Gamma^\al\equiv -g^{\gm\bt} \Gamma^\al_{\gm\bt}, \ee
where $\Gamma^\al_{\gm\bt}$ are the usual Christoffel symbols.

It was realized by Friedrich \cite{Friedrich85} and also by Garfinkle
\cite{Garfinkle2001}, that it is possible to eliminate the mixed
derivatives in the principal part of the Einstein equations while
largely recovering the coordinate freedom than is lost by choosing
the harmonic gauge.
Instead of (\ref{Harmonic}), one requires that that the coordinates
satisfy
\be
\label{GH_coords}
\Box x^\al = H^\al, \ee
where $H_\al\equiv g_{\alpha\beta}H^\beta$ are arbitrary ``gauge
source functions'' \footnote{In a slight abuse of notation and
  terminology we will refer to both $H_\alpha$ and $H^\alpha$ as
  ``the'' gauge source functions.  } which are to be viewed as
specified quantities.  One then defines the GH constraint
\be
\label{Ca}
C^\al \equiv H^\al - \Box x^\al, \ee
which clearly must vanish provided~(\ref{box_x}) holds, and then
modifies the Einstein equations as follows:
\be
\label{eqH0}
R_{\mu\nu} - C_{(\mu;\nu)} = \bar{T}_{\mu\nu}. \ee
This last equation can be written more explicitly as
\be \label{eqH} -\half g^{\al\bt} g_{\mu\nu,\al\bt} -
{g^{\al\bt}}_{(,\mu} g_{\nu)\bt,\al} - H_{(\mu,\nu)} +
H_\bt\Gamma^\bt_{\mu\nu} -\Gamma^\al_{\nu\bt} \Gamma^\bt_{\mu\al}=
\bar{T}_{\mu\nu}. \ee
Now, provided that the $H_\al$ are functions of the coordinates and
the metric only, but not of the metric derivatives---namely
$H_\al=H_\al(x,g)$---the field equations~(\ref{eqH}) form a manifestly
hyperbolic system. We reemphasize that the source functions $H_\al$
are arbitrary at this stage and that their specification is equivalent
to choosing the coordinate system for the spacetime under
consideration (``fixing the gauge'').  Determining an effective
prescription for the source functions is thus crucial for the efficacy
of the GH approach, and several strategies for fixing the $H_\al$ are
discussed in the next section.

Having prescribed the coordinates we integrate the equations forward
in time.  Consistency of the scheme requires that the GH constraint
(\ref{Ca}) be preserved in time.  The contracted Binachi identities
guarantee that this is indeed the case, since, using those identities,
one can show \cite{FP1, Lindblometal1} that $C^\al$ itself satisfies a
wave equation,
\be
\label{C_eq}
\Box C^\al +{R^\al}_\nu \, C^\nu =0. \ee
Thus, assuming that the evolution is generated from an initial
hypersurface on which $C^\al=\pa_t C^\al=0$, (\ref{C_eq}) guarantees
that $C^\al=0$ for all future (or past) times.

Although the GH constraint is preserved at the continuum level, in
numerical calculations, where equations are discretized on a mesh with
some characteristic mesh scale, $h$, the constraint cannot be expected
to hold exactly.  More troublingly, experience shows that numerical
solutions of~(\ref{eqH})---particularly in strong field cases, such as
those involving black holes---can admit ``constraint violating
modes'', with the result that the desired continuum solution is {\em
  not} obtained in the limit $h\to0$.  Fortunately, an effective way
of preventing the development of such modes in numerical calculations
exists: one adds terms to the field equations that are explicitly
designed to damp constraint violations (see e.g.~\cite{KST}).  In our
implementation we follow Pretorius \cite{FP1,FP3} by adding constraint
damping terms in a fashion inspired by studies of the so-called
$\gm$-systems \cite{gm_sys,Gundlachetal}.  The modified equations take
the form
\bea \label{Eqs_constrdamp} -\half g^{\al\bt} g_{\mu\nu,\al\bt} &-&
{g^{\al\bt}}_{(,\mu} g_{\nu)\bt,\al} - H_{(\mu,\nu)} +
H_\bt\Gamma^\bt_{\mu\nu} -\Gamma^\al_{\nu\bt} \Gamma^\bt_{\mu\al} -
\non & -& \kappa \( n_{(\mu}\cC_{\nu)} -\half g_{\mu\nu} \, n^\bt \,
\cC_\bt \) = \bar{T}_{\mu\nu}. \eea
Here, $n^\al$ is the future-directed, unit time-like vector normal to
the $t={\rm const.}$ hypersurfaces, which can be written as
\be
\label{na}
n_\al \equiv -\(1/\sqrt{-g^{tt}}\) \pa_\al t ,
% = \(-(-g^{tt})^{-1/2},0 \).
\ee
and $\kappa$ is an adjustable parameter that controls the damping
timescale.  Specifically, as discussed in~\cite{Gundlachetal}, small
constraint perturbations about a fixed background decay exponentially
with a characteristic timescale of order $\kappa$.  We note that the
constraint damping term contains only first derivatives of the metric
and hence does not affect the principal (hyperbolic) part of the
equations.
%%%%%%%%%%%%%%%%%%%%%%%%%%%%%%%%%%%%%%%%%%%%%%%%%%%%%%%%%%%%%%%%%
\section{Coordinate conditions}
\label{sec_gauge_conditions}
%%%%%%%%%%%%%%%%%%%%%%%%%%%%%%%%%%%%%%%%%%%%%%%%%%%%%%%%%%%%%%%%%

As we have already mentioned, fixing the coordinates in the GH
approach amounts to specifying the source functions $H_\al$.  In this
regard, it is instructive to examine the relationship between the
$H_\al$ and the lapse function and shift vector that appear in the
ADM, or space-plus-time,
 formulation of general relativity.  
We recall that in the ADM formalism the line element can be written as
\be
\label{ADM_metric}
ds^2=-\al^2 dt^2+\gm_{ij}\(dx^i+\bt^i dt\)\(dx^j+\bt^j dt\) ,  \ee
where $\al$ is the lapse function, $\bt^i$ is the shift vector, and
$\gm_{ij}$ is the spatial metric of the $t={\rm const.}$ hypersurfaces.
Using this form of the spacetime metric in (\ref{GH_coords}) yields
\bea
\label{albt-H}
\pa_t \al -\bt^k \pa_k \al &=& -\al\(H_n +\al K\),\non \pa_t \bt^i
-\bt^k \pa_k \bt^i &=& \al \gm^{ij} \[\al \( H_j+
{^{(D-1)}}\Gamma_{jkl}\gm^{kl} \) -\pa_j \al\] , \eea
where $H_n \equiv n^\mu H_\mu = (H_t-\bt^i H_i)/\al$ is the normal
component of the source function $H_\mu$, $K$ is the trace of the
extrinsic curvature tensor of the $t={\rm const.}$ slices, and the
$^{(D-1)}\Gamma_{jkl}$ are Christoffel symbols associated with the
spatial $\gm_{ij}$.  Bearing in mind that the temporal component of
the source function is thus determined by $H_t= \al\, H_n +\bt^i H_i$,
these last equations clearly exhibit the connection between the gauge
source functions and the time evolution of the lapse and shift.

In his groundbreaking application of the GH approach~\cite{FP2,FP3},
Pretorius used insight derived from considering this relationship
between the $H_\al$ and the ADM kinematic variables to devise a
methodology that generates effective gauge source functions for the
problem of binary black hole collisions.  His strategy elevates the
status of the $H_\al$ to independent {\it dynamical} variables that
satisfy time-dependent partial differential equations.  Crucially, the
evolution equations for the $H_\alpha$ are designed so that the lapse
and shift which (implicitly) result from the time development have
certain desirable properties.  For example, the equation for $H_t$ is
tailored in an attempt to keep the value of the lapse function of
order unity everywhere---including near the surfaces of the black
holes---during the evolution.

One specific prescription for achieving this type of control evolves
the gauge source functions according to
\bea
\label{FP_gauge}
\Box H_t& =& -\xi_1 \,\frac{\al-\al_0}{\al^q} + \xi_2 \, H_{t,\mu}
n^\mu ,\non H_i &=&0, \eea
where $\Box$ is the covariant wave operator, and $\al_0, \xi_1, \xi_2$
and $q$ are adjustable constants\footnote{Sometimes it is convenient
  to assume that $\xi_1$ and $\xi_2$ are given functions of space and
  time rather than mere constants.  For example, one might require that
  the gauge driver is switched on gradually in time, or that it be
  active only in certain regions, e.g.~in the vicinity of a black
  hole, and that its effect vanish asymptotically, so that pure 
  harmonic coordinates are recovered at large distances. }.  Thus the temporal source
function satisfies a wave equation similar to those that govern the
metric components in the system (\ref{Eqs_constrdamp}).  The first
term on the right-hand-side of~(\ref{FP_gauge}) is designed to
``drive'' $H_t$ to a value that results in a lapse that is
approximately $\al_0$.  The second, ``frictional'' term tends to
confine $H_t$ to this value.  For the case of the spatial coordinates,
Pretorius found that the simplest choice of spatially harmonic
gauge---$H_i=0$---was sufficient in simulations of binary black hole
collisions.  Importantly, the choice (\ref{FP_gauge}) ensures that the
hyperbolicity of the combined evolution system is preserved.  A slight
generalization of this technique was considered in \cite{Scheel_etal}
where instead of using $H_i=0$, the spatial components of the source
functions are evolved according to
\be
\label{FP_gauge_m}
\Box H_i = -\xi_3 \,\frac{\bt_i}{\al^2} + \xi_2 \, H_{i,\mu}n^\mu \ee
where $\xi_3$ is an additional parameter.

One possible problem with the specific driver approach outlined above
is that the coordinates that result do not correspond to those
produced by any of the more familiar coordinate conditions typically
used in numerical relativity.  Recently, Lindblom {\em et al}
\cite{Lindblometal2} proposed driver conditions that are crafted so
that the source functions that result imply {\em particular}
conditions on the corresponding lapse and shift.  We now proceed to a
review of this interesting and promising approach.

We begin by observing that many traditional coordinate conditions of
numerical relativity can be written as $F_\al=F_\al(x,g,\pa g)$ where
the $F_\al$ are to be viewed as ``effective'' gauge source functions
which could be computed, for example, were the entire spacetime in
hand.  Within the GH approach, enforcing such a condition
algebraically by simply setting $H_\al=F_\al$ will generally destroy
the hyperbolicity of the system, since the $ H_{(\mu;\nu)}$ terms in
(\ref{eqH}) will generically give rise to mixed second derivatives of
the metric.  Lindblom {\em et al} circumvent this difficulty by
generalizing (\ref{FP_gauge}) to
\be
\label{H_driver1}
\O H_\al = Q_\al(x,g,\pa g, H, \pa H), \ee
where $\O$ is a second order hyperbolic operator and $Q_\al$ is chosen
so that the source functions evolve towards the concrete
$F_\al=F_\al(x,g,\pa g)$ that define the desired gauge.  The combined
system (\ref{Eqs_constrdamp}) and (\ref{H_driver1}) will remain
hyperbolic provided the $Q_\al$ depend on at most first derivatives of
the fields. In analogy with (\ref{FP_gauge}) the authors of
\cite{Lindblometal2} choose
\be
\label{Q_al1}
Q_\al = \mu_1^2 \,\(H_\al-F_\al\)+2\mu_2 \, \pa_t H_\al +\eta\,W_\al,
\ee
where $\mu_1, \mu_2$ and $\eta$ are adjustable parameters, and $W_\al
$ is assumed to satisfy
\be
\label{W_al}
\pa_t W_\al + \eta \, W_\al = \hat {\O} H_\al, \ee
where $ \hat {\O}$ is the part of $\O$ that contains only spatial
derivatives. When the spacetime is stationary, time-derivatives vanish
and equations (\ref{H_driver1}) and (\ref{Q_al1}) then imply
$H_\al=F_\al$.  Notice that without the introduction of the auxiliary
fields, $W_\al$, this property could not be attained for general,
position dependent gauges \cite{Lindblometal2}.

In order to implement this method for a specific desired gauge choice
one must first compute the corresponding target source functions,
$F_\al$.  Here we focus on gauges of the schematic form $G_\al(x, g,
\pa g)=0$ for which one can choose \cite{Lindblometal2}
\be
\label{F_al1}
F_\al=-\Gamma_\al -q\, G_\al, \ee
where $q$ is a tunable parameter. In the GH formalism,
$H_\al=-\Gamma_\al$, and (\ref{F_al1}) then implies $H_\al-F_\al=q\,
G_\al$.  This demonstrates that when the GH constraint is satisfied,
$H_\al$ is driven to $F_\al$ if $G_\al$ is driven to zero. We next
discuss several specific coordinate choices that are explored in this
paper.

%%%%%%%%%%%%%%%%%%%%%%%%%%%%%%%%%%%%%%%%%%%%
\subsection{Slicing conditions}
\label{sec_slicing}
%%%%%%%%%%%%%%%%%%%%%%%%%%%%%%%%%%%%%%%%%%%%
For the particular choices of the slicing conditions that we use in
this paper, it is more convenient to calculate the normal component of
the target source functions, $F_n\equiv n^\mu F_\mu=(F_t-\bt^i
F_i)/\al$, than the temporal component, $F_t$, itself 
(see (\ref{albt-H})).
Once this is done, then
in conjunction with the shift conditions that fix $F_i$, the temporal
component can be easily computed via $F_t=\al F_n +\bt^i F_i$.
\begin{itemize}

\item Constant curvature slicing, $K=K_0$. Here we assume that the
  trace, $K(g,\pa g)$, of the extrinsic curvature of the spatial
  slices is constant. When $K_0=0$, we have the famous maximal slicing
  condition \cite{SmarrYork} whose significant popularity in numerical
  calculations is due in large part to the strong singularity-avoiding
  property exhibited by the resulting constant-time surfaces (see
  Sec.~\ref{sec_results}).  The constant curvature foliation can be
  written as $G_n=0$, where
  \be
  \label{Gn_K0}
  G_n=K_0-K=K_0+\nabla^\al n_\al \ee

\item Bona-Masso slicing \cite{BonaMasso}. This condition can be
  written as
  \be
  \label{Gn_BM}
  G_n=(\pa_t \al -\bt^i \pa_i \al) +\al^2 f(\al) \(K-K_0\), \ee
  where $f(\al)$ is an arbitrary function of the
  lapse.\footnote{Sometimes the geometric derivative $\pa_n \al \equiv
    (\pa_t \al -\bt^i \pa_i \al)/\al $ is replaced with the partial
    time derivative $\pa_t \al$.} The choice $f(\al)=2/\al$
  corresponds to the popular $1+\log$ slicing.

\end{itemize}

In terms of implementing these slicing conditions, we note that
(\ref{F_al1}) implies
\be
\label{Fn}
F_n = -\al^{-1}\(\Gamma_t-\bt^i \Gamma_i\)-q_n\, G_n, \ee
where $q_n$ is a parameter, and that the kinematic quantities such as the
lapse and shift which appear in various formulae above can always be
written in terms of the fundamental dynamical variables of the scheme
(i.e.~the metric components and their first derivatives).

%%%%%%%%%%%%%%%%%%%%%%%%%%%%%%%%%%%%%%%%%%%%
\subsection{Shift conditions}
\label{sec_shift-cond}
%%%%%%%%%%%%%%%%%%%%%%%%%%%%%%%%%%%%%%%%%%%%

An important class of shift conditions which is often used in
numerical relativity employs versions of the so-called $\Gamma$-driver
\cite{ABDKPST}.  In this approach, one first introduces the
conformally rescaled spatial metric, $\tilde{\gm}_{ij} = \gm^\sigma
\gm_{ij}$, with $\gm \equiv \det \gm_{ij} $ and $\sigma$ an arbitrary
parameter, then computes the contracted Christoffel symbols,
\be
\label{Gam_conf}
^{(D-1)}\tilde{\Gamma}^i = ^{(D-1)}\tilde{\Gamma}^i_{kj}
\tilde{\gm}^{kj} =- \gm^{-\sigma}\[\frac{1+\sigma \,(D-3)}{2} \gm^{ij}
\pa_j \log \gm + \gm ^i_j \pa_k \gm^{kj}\], \ee
and imposes certain conditions on their dynamics.  The $\Gamma$-driver
strategy is related to the minimal distortion condition
\cite{SmarrYork,BaumgarteShapiro} which is designed to minimize the
time variation of $\tilde{\gm}_{ij}$ (see e.g.~\cite{ABDKPST}).

\begin{itemize}

\item $\Gamma$-freezing. Here one requires
  \be
  \label{Gamma_freezing}
  \pa_t\,^{(D-1)}\tilde{\Gamma}^i = 0, \ee
  which implies that during the evolution $ ^{(D-1)}\tilde{\Gamma}^i$
  is fixed,
  $^{(D-1)}\tilde{\Gamma}^i=^{(D-1)}\tilde{\Gamma}^i|_{t=0}$.
  Following \cite{Lindblometal2} we attempt to evolve to this choice
  by choosing
  \be
  \label{Gi_Gamma-freezing}
  G_i=
  \tilde{\gm}_{ij}\(^{(D-1)}\tilde{\Gamma}^j(0)-^{(D-1)}\tilde{\Gamma}^j\).
  \ee
\item $\Gamma$-driver.  Again following \cite{Lindblometal2} we write
  the driver condition as
  \bea
  \label{Gamma-driver-1_dtbt}
  \pa_t \bt^i& =& \nu \[ ^{(D-1)}\tilde{\Gamma}^i - \eta_2 \,B^i \],
  \\
  \label{Gamma-driver-1_dtB}
  \pa_t B^i &+&\eta_2\, B^i = ^{(D-1)}\tilde{\Gamma}^i, \eea
  where $\nu$ and $\eta$ are adjustable parameters. Then one can
  choose
  \be
  \label{Gi-Gamma-driver-1}
  G_i= \gm_{ij}\(\pa_t \bt^j - \nu\,^{(D-1)}\tilde{\Gamma}^j +\nu
  \eta_2 \,B^j \), \ee
  The auxiliary variable $B^i$ is evolved using
  (\ref{Gamma-driver-1_dtB}) and it is important to note that adding
  this equation to the scheme does not destroy the hyperbolicity of
  the combined evolution system \cite{Lindblometal2}.

  We have also experimented with a geometric version of the driver
  where the partial time derivative $\pa_t$ in
  (\ref{Gamma-driver-1_dtB}) is replaced with the covariant derivative
  $n^\mu \nabla_\mu \equiv (\pa_t -\bt^k \pa_k)/\al $.

\end{itemize}

Implementation of the above shift conditions is effected by setting
the corresponding spatial target source function defined by
(\ref{F_al1}) according to
\be
\label{Fi}
F_i = -\Gamma_i -q_i\, G_i, \ee
where $q_i$ is an adjustable parameter.

%%%%%%%%%%%%%%%%%%%%%%%%%%%%%%%%%%%%%%%%%%%%%%%%%%%%%%%%%%%
\section{Spherically-symmetric reduction}
\label{sec_formulation}
%%%%%%%%%%%%%%%%%%%%%%%%%%%%%%%%%%%%%%%%%%%%%%%%%%%%%%%%%%%%
Having described the basics of the GH formalism, we now specialize to
spherically symmetric spacetimes. We consider a $D$-dimensional
spacetime with $SO(D-2)$ rotational symmetry, and write the
$D$-dimensional line element in the form
\be
\label{SO_D-2_metric}
ds^2= g^{(D)}_{\mu\nu} dx^\mu dx^\nu = g^{(D)}_{ab} dx^a dx^b + e^{2\,
  \hat {S}} d\Omega_{n}^2. \ee
Here $ d\Omega_{n}^2$ is the metric on a unit n-sphere, $n\equiv D-2$, $a,b = \{t,r\}$, and the metric $g^{(D)}_{a b}$
and scalar $\hat{S}$ are functions of $t$ and the radial coordinate,
$r$, alone.

Although we will later specialize to the case of a real, massless 
scalar field, for generality we first adopt as a matter source 
 minimally coupled complex scalar field, $\Phi$, with a
potential $V(|\Phi|)$.  The action that describes
the system can be written as
\be
\label{action}
S = \int \sqrt{-g^{(D)}}\( R^{(D)} -\pa_a \Phi\, \pa^a \Phi^* - 2\,
V(|\Phi|)\)dx^D.  \ee
By varying the action with respect to the fields one gets the Einstein
equations (\ref{Eeqs}) with the energy-momentum tensor
$\bar{T}_{\mu\nu} = \half\(\pa_\mu \Phi \, \pa_\mu \Phi^*+ \pa_\mu
\Phi^* \, \pa_\mu \Phi\)+ 2/(D-2) g_{\mu\nu}^{(D)} \, V$, as well as
the general relativistic Klein-Gordon equation for the scalar field.
Specifically, the GH transformation of the Einstein
equations as given by~(\ref{eqH0}) reads
\bea &&R^{(D)}_{ab} - C_{(a;b)} = \half\(\pa_a \Phi \, \pa_b \Phi^*+
\pa_a \Phi^* \,
\pa_b \Phi\) + \frac{2}{D-2} g_{ab}^{(D)}\, V, \label{Eqgab} \\
&&R^{(D)}_{\theta_i\theta_i} - C_{(\th_i;\th_i)} = \frac{2}{D-2}  g_{\theta_i\theta_i}^{(D)}\,V,  \label{EqS}\\
&&\Box \Phi = \pa V/\pa \Phi^*\label{eqPhi}, \eea
where $ R^{(D)}_{\mu\nu} $ is the $D$-dimensional Ricci tensor and
$\theta_i$ are the angular coordinates.  In spherical symmetry it
suffices to use any specific angular component of the Ricci tensor,
and for convenience we use $R^{(D)}_{\theta_1\theta_1}$ where
$\theta_1$ is defined by $d\Omega_n^2=d\theta_1^2+\sin^2 \theta_1
d\Omega_{n-1}^2$.

The form of the metric (\ref{SO_D-2_metric}) is not yet optimal for
use in numerical computations.  In this paper we are mostly interested
in asymptotically flat solutions and thus the following section
describes a more natural ansatz for use in that instance.

%%%%%%%%%%%%%%%%%%%%%%%%%%%%%%%%%%%%%%%%%%%%%%%%%%
\subsection{Spatial asymptotics}
\label{sec_asympt}
%%%%%%%%%%%%%%%%%%%%%%%%%%%%%%%%%%%%%%%%%%%%%%%%%%

In spherical coordinates, flat spacetime can be written as
\be
\label{flat1}
ds^2= - dt^2 + dr^2 +r^2 \,d\Omega_n^2.  \ee
It follows from (\ref{SO_D-2_metric}) that asymptotically $g_{ab}
\rightarrow \eta_{ab}$, where $\eta_{ab}$ is a Minkowski metric, and
$\hat{S} \rightarrow \log \, r$, (i.e.~$\hat{ S}$ diverges at spatial
infinity).  Since this divergence complicates the numerical
implementation of boundary conditions, we introduce a new function,
$S$, defined by  $S=\hat{S}- \log \, r $, which is regular everywhere.  
We then adopt the following,
more regular form for the line element in the asymptotically flat
case:
\be
\label{flat_ds}
ds^2=g_{ab} dx^a dx^b +r^2\, e^{2\,S} d\Omega^2_n. \ee
In spherical coordinates, the source function derived from
(\ref{GH_coords}) does not vanish even in flat spacetime where it
becomes 
\be
H^{\rm Mink}_\mu=-\Gamma^{\rm Mink}_\mu=(0, n/r, (n-1)\,\cot
\theta_1,(n-2)\,\cot\theta_2,\dots,\cot \theta_{n-1},0) .
\ee
Since near the origin spacetime is locally flat, the radial component of the
source function is generically singular at $r=0$, diverging as $n/r$.
To regularize this radial component, we thus subtract the singular
background contribution by transforming $H_\al \to H_\al +\delta^r_\al
H_r^{\rm Mink}$, and use the functions $H_t$ and $H_r$ defined by
\be
\label{H_mink}
H_\al=\(H_t(t,r),H_r(t,r)+n/r,(n-1)\cot
\theta_1,(n-2)\,\cot\theta_2,\dots,\cot \theta_{n-1},0\). \ee
in our formulae.

With the line-element (\ref{flat_ds}) and the source functions
(\ref{H_mink}), the asymptotic behavior of the fields is simply
\be
\label{asymp_bc}
g_{ab} \rightarrow \eta_{ab} ,~~~ S \rightarrow 0, ~~~\phi \rightarrow
0 , ~~~ H_t \rightarrow 0, ~~~H_r \rightarrow 0 \ee
In App.~\ref{sec_asympt_AdS} we also analyze the asymptotically
AdS spacetime, which is described in our model (\ref{action}) for the
case that the scalar field potential satisfies $V(0)\rightarrow
\Lambda<0$.
%%%%%%%%%%%%%%%%%%%%%%%%%%%%%%%%%%%%%%%%%%%%%%%%%%%%%%%%%%%%%%%%%
\subsection{Center of symmetry, $r=0$}
\label{sec_axis_bc}
%%%%%%%%%%%%%%%%%%%%%%%%%%%%%%%%%%%%%%%%%%%%%%%%%%%%%%%%%%%%%%%%%

Invariance of the line element (\ref{flat_ds}) under the reflection
$r\rightarrow -r$ in spherical symmetry implies that $g_{tr}$ is an
odd function of $r$, while $g_{tt}, g_{rr}, S$ and $\Phi$ are even in
$r$.  Additionally, the GH constraint (\ref{GH_coords}) implies that
the source functions $H_r$, regularized via (\ref{H_mink}), and $H_t$
are odd and even in $r$, respectively.

Moreover, the requirement that the surface area of an $n$-sphere must
vanish at the origin\footnote{that is, that the radial and areal
  coordinates coincide at the origin, to avoid a conical singularity
  there.} implies $g_{rr}(t,0)=e^{2\,S(t,0)} $.  We note that this is
an extra condition on $S$, which thus has to satisfy both this
relation, as well as the constraint that it have vanishing radial
derivative at $r=0$---specifically that $g_{rr}-e^{2\,S} = O(r^2)$.
Therefore, at $r=0$ we essentially have three conditions on the two
fields $S$ and $g_{rr}$.  In the continuum, and given regular initial
data, the evolution equations will preserve regularity: however, in a
numerical code that solves the equations discretized on a lattice,
this will be true only up to discretization errors.  As a general
rule-of-thumb, the number of boundary conditions should be equal to
the number of evolved variables in order to avoid regularity problems
and divergences of a numerical implementation.

An elegant way to deal with this regularity issue involves definition
of a new variable, $\lambda$:~\footnote{We note that a similar variable
was introduced in 
\cite{lambda_ref}, also for the purpose of regularization.}
\be
\label{lambda_var}
\lambda \equiv \frac{g_{rr} -e^{2\, S}}{r}.  \ee
At the origin one then has $\lam \sim O(r)$.  Therefore, after
changing variables from $S$ to $\lam$ by using $ S=(1/2) \log(g_{rr}-
r\, \lam)$ in all equations, and imposing $\lam(t,0)=0$ at the origin,
one ends up with a system where there is no over-constraining due
to the demand of regularity at $r=0$. In addition, we note that at
spatial infinity we have $\lam=0$, and that the hyperbolicity of the
GH system is not affected by the change of variables.

However, as described in detail in Sec.~\ref{sec_coords_bc}, 
we were able to implement a more straightforward regularization
method that maintains
$S$ as a fundamental dynamical variable, and thus opted to use that
approach in our current calculations.

\subsection{The equations}
\label{sec_eqs}

With the metric ansatz (\ref{flat_ds}) and the regularized source
function
(\ref{H_mink}), equations (\ref{Eqgab})--(\ref{eqPhi}) become 5
equations for the 5 variables, $\gtt, \gtr,\grr, S$ and $\Phi$, that
schematically can be written as\footnote{Using $\lambda$ instead of
  $S$ does not change this structure since the equation that governs
  $\lambda$ is a linear combination of the equations that govern $S$
  and $g_{rr}$.}
\bea && -\half g^{cd} g_{ab,cd} + \dots = \half\( \pa_a \Phi \, \pa_b
\Phi^*+\pa_a \Phi^* \, \pa_b \Phi\)
+ \frac{2}{D-2} g_{ab} V, \label{EqHgab} \\
&&  g^{cd} S_{,cd} + \dots=  -\frac{2}{D-2}\, V  , \label{EqHS} \\
&& g^{cd} \Phi_{,cd} + \dots = \pa V/\pa\Phi^*\label{EqHPhi}. \eea
Here ellipses denote terms that may contain the metric and/or the
source functions, as well as their first derivatives in various
combinations (see App.~\ref{sec_SS_eqs} for the explicit set of
equations in the four-dimensional case).  These equations are to be
evolved forward in time starting from the initial ($t=0$) time slice,
where values for the fields and their first time derivatives must be
prescribed.

%%%%%%%%%%%%%%%%%%%%%%%%%%%%%%%%%%%%%%%%%%%%%%%%%%%%%%%%%%%%%%%%%
\subsection{Coordinate choices}
\label{sec_SS_gauges}
%%%%%%%%%%%%%%%%%%%%%%%%%%%%%%%%%%%%%%%%%%%%%%%%%%%%%%%%%%%%%%%%%
Here we adapt the prescriptions for choosing the gauge functions
($H_t$ and $H_r$) that were described in Sec.~\ref{sec_gauge_conditions}, to the case of spherical symmetry.  We
again note that the radial source function is singular at the origin
in spherical symmetry, and that we thus regularize it
via~(\ref{H_mink}).  Since this regularization involves subtracting
the flat-spacetime singular part from $H_r$, any specific coordinate
conditions discussed here are thus defined relative to spherical
Minkowski spacetime.

For the case of the gauge condition~(\ref{FP_gauge}) inspired by
Pretorius' original work, we have
\bea
\label{FP_gauge_SS}
\Box H_t& =& -\xi_1 \,\frac{\al-\al_0}{\al^q} + \xi_2 \, \(\pa_t H_t
-\bt \,\pa_r H_t\)/\al, \\ H_r &=&0. \nonumber \eea
Similarly for the modification of the above proposed in
\cite{Scheel_etal}, we have (using~(\ref{FP_gauge_m}))
\bea
\label{FP_gauge_SS_m}
\Box H_t& =& -\xi_1 \,\frac{\al-\al_0}{\al^q} + \xi_2 \, \(\pa_t H_t
-\bt \,\pa_r H_t\)/\al,\\ \Box H_r &=& -\xi_3 \,\frac{\bt}{\al^2} +
\xi_2 \, \(\pa_t H_r -\bt\, \pa_r H_r\)/\al. \nonumber \eea
In the above equations $\Box$ is the regularized scalar wave operator in spherical
symmetry, given by
\be
\label{Lind_scalar}
\Box\,H_\al = g^{\mu\nu} \pa_\mu\pa_\nu H_\al - \(\Gamma^\nu +
g^{rr}\,\frac{n}{r} \delta^\nu_r \delta^\al_r\)\pa_\nu\,H_\al.  \ee
Turning now to the case of the gauge drivers introduced by Lindblom
{\em et al}, we note that the operator in (\ref{H_mink}) is
essentially the vector d'Alambertian\footnote{$H_a$ does not transform
  as a vector under gauge transformations, so the equation should be
  understood as written in particular global coordinates
  \cite{Lindblometal2}; in the current case, these are our spherical coordinates.}
\cite{Lindblometal2}
\be
\label{Lind_wave1}
\O\,H_\al = g^{\mu\nu} \pa_\mu\pa_\nu H_\al -\Gamma^\nu \pa_\nu H_\al-
2\, g^{\mu\nu} \Gamma^{\bt}_{\nu\al} \pa_\mu H_\bt +
\(R_\al^\bt-\pa_\al \Gamma^\bt\)\,H_\bt. \ee
In order to avoid having second-derivatives of the metric, the Ricci
tensor in the last term should be thought of as being
determined by matter
sources and replaced with $\bar T_\al^\bt$, in accordance with the
Einstein equations.
In addition, using the GH constraint $H_\al = - \Gamma_\al$, the term
$-\pa_\al \Gamma^\bt\,H_\bt$ is replaced with $-\pa_\al
H^\bt\,\Gamma_\bt $.  Finally, we regularize the operator by
subtracting the irregular contributions that appear in the flat
spacetime limit.  After these manipulations we arrive at
\be
\label{Lind_wave2}
\O\,H_\al = g^{\mu\nu} \pa_\mu\pa_\nu H_\al - \(\Gamma^\nu +
g^{rr}\,\frac{n}{r} \delta^\nu_r \delta^\al_r\)\pa_\nu\,H_\al- 2\,
g^{\mu\nu} \Gamma^{\bt}_{\nu\al} \pa_\mu H_\bt -
\(\bar{T}_\al^\bt+\pa_\al H^\bt\)\(\Gamma_\bt + \frac{n}{r}
\delta_{\al_r}\), \ee
where $\delta_\mu^\nu$ is a Kronecker delta, and there is no
summation over the index $\al$.

The target source function, $F_n$, is determined by (\ref{Gn_K0}) or
(\ref{Gn_BM}), and by (\ref{Fn}).  The lapse and shift are given in
terms of the metric components,
\bea
\label{albt-gab}
\al &=& \sqrt{-\gtt + \gtr^2/\grr}, \\
\bt &=& \gtr/\grr, \nonumber \eea
as is the trace of the extrinsic curvature (see (\ref{trK}) for the
explicit form).

Our shift conditions involve the contracted conformal Christoffel
symbols, $\tilde{\Gamma}_i$, defined by (\ref{Gam_conf}), and in
spherical symmetry the only non-trivial component is
$^{(D-1)}\tilde{\Gamma}_r$ given by
\be
\label{ConfGammaSS}
^{(D-1)}\tilde{\Gamma}_r=-n\,\(1+(n-1)\,\sigma\)\,S' +
\frac{1-\sigma\,(n-1)}{2} \frac{g_{rr}'}{g_{rr}}. \ee
Here $()' \equiv \pa_r$, and we have used the fact that
$\gm_{rr}=g_{rr}$.  Once again, in order to obtain a regular
expression we have subtracted the flat-spacetime term,
$^{(D-1)}\tilde{\Gamma}_r^{\rm Mink} = -n(1+(n-1)\sigma)/r$, which is
singular at the origin.

The target function for the $\Gamma$-freezing condition
(\ref{Gi_Gamma-freezing}) takes the form
\be
\label{Fi_Gamma-freezing_SS}
F_r =-\hat\Gamma_r -
q_s\[^{(D-1)}\tilde{\Gamma}_r(0,r)\,\(\frac{\grr}{\grr(0,r)}\)^{\sigma+1}\,e^{2\,n\,\sigma\,[S-S(0,r)]}
- ^{(D-1)}\tilde{\Gamma}_r\], \ee
where $\hat\Gamma_r \equiv \Gamma_r + n/r$ is the $D$-dimensional
connection which has also been regularized via subtraction of an
irregular flat-spacetime term. The explicit expression for $\Gamma_r$
is given in (\ref{Gammas}).

For the case of the $\Gamma$-driver condition (\ref{Fi}) in spherical
symmetry, the target source function is
\be
\label{Fi_Gamma-driver-1_SS}
F_r=-\hat \Gamma_r - q_s \[ \grr\,\dot \bt^r -
^{(D-1)}\tilde{\Gamma}_r\,\nu\, (\grr e^{2\, n\, S})^{-\sigma} +
\nu\,\eta_2 \, \grr\,B\] , \ee
where an over-dot denotes partial differentiation with respect to $t$.
The auxiliary field $B$ is evolved using
\be
\label{eqB_SS}
\dot B +\eta_2 B = ^{(D-1)}\tilde{\Gamma}_r\(\,e^{2\, n\,
  S}\,\grr\)^{-\sigma}/\grr.  \ee
%

%%%%%%%%%%%%%%%%%%%%%%%%%%%%%%%%%%%%%%%%%%%%%%%%%%%%
\subsection{Initial data}
\label{sec_initdata}
%%%%%%%%%%%%%%%%%%%%%%%%%%%%%%%%%%%%%%%%%%%%%%%%%%%%
We now consider specification of initial data, which as stated
previously, are values for the fields and their first time derivatives
at $t=0$.  For simplicity (and without much loss of generality), we
restrict attention to time-symmetric initial conditions.

Given the assumption of time symmetry at $t=0$, initial data for the
scalar field reduces to the specification of $\Phi(0,r)$, which we
take to have the form of a Gaussian,
\be
\label{Phi0}
\Phi(0,r)=\Phi_0\, e^{-(r-r_0)^2/\Delta^2}, \ee
where $\Phi_0$, $r_0$ and $\Delta$ are adjustable parameters.

The momentum constraint is trivially satisfied for time-symmetric
initial data, and writing the initial metric as
\be
\label{TS_metric}
ds^2=-\al^2 dt^2 +\psi^4(dr^2+r^2 d\Omega_n^2),  \ee
the Hamiltonian constraint becomes a non-linear ordinary differential
equation for $\psi(0,r)$,
\be
\label{Hc}
\psi''+{n \over r} \psi' +(n-2){\psi'^2 \over \psi} + \frac{1}{2
  n}\(\half \Phi'\,{\Phi^*}'+ \psi^4\, V\)\, \psi =0.  \ee
This equation is solved using the boundary conditions
$\psi'(0,r)|_{r=0}=0$ and $\psi(0,r)|_{r \rightarrow \infty}=1$, and
then once $\psi$ has been determined, the metric components are
initialized via
\bea \label{TS_init} g_{rr} &=& \psi^4, \non
S&=&2\, \log \psi, \\
g_{tr} &=& \bt^r=\lam=0 . \nonumber \eea
For time-symmetric initial data we require that all first time
derivatives of the metric components vanish.

We next determine the initial conditions for the lapse and the
variables used in the gauge drivers. We begin by setting
$H_t(0,r)=H_r(0,r)=0$.  Using
\be
\label{TS_Ha}
H_r(0,r) =
-\hat\Gamma_r(0,r)=\frac{\al'}{\al}+2(n-1)\frac{\psi'}{\psi}, \ee we
obtain an equation relating $\al(0,r)$ to the initial value of $H_r$.
With our choice, $H_r(0,r)=0$, this equation can be integrated to
yield
\be
\label{al_0}
\al(0,r)=\psi(0,r)^{-2(n-1)}.  \ee
Next we require that the target coordinate conditions are initially
satisfied, namely that
$F_\al(0,r)=G_\al(0,r)=H_\al(0,r)=0$. We note that since 
time-symmetry implies $K(0,r)=K_0=0$, the normal component of the
gauge function for the constant curvature foliation vanishes,
$G_n(0,r)=q_n K_0=0$, as it does for the Bona-Masso slicing,
$G_n(0,r)=-q_n\,\al(0,r)^2 f(\al(0,r))\,K_0=0$.  The $\Gamma$-freezing
condition (\ref{Fi_Gamma-freezing_SS}) obviously satisfies
$G_i(0,r)=0$, while requiring this for the $\Gamma$-driver condition
(\ref{Fi_Gamma-driver-1_SS}) will set the initial value of the
auxiliary field $B$\footnote{Note that for time-symmetric initial
  conditions this consistently coincides with the values of $B(0,r)$ found from
  (\ref{eqB_SS}).},
\be B(0,r)=^{(D-1)}\tilde{\Gamma}_r \,
e^{-2\,n\,S\,\sigma}\grr^{-\sigma}/\eta_2|_{t=0}. \ee
Here the initial value for the radial component of the contracted
conformal Christoffel symbol $\tilde{\Gamma}_r(0,r)$, defined by
(\ref{ConfGammaSS}), is found using the relations~(\ref{TS_init}):
\be
\label{ConfGammaSS_init}
^{(D-1)}\tilde{\Gamma}_r(0,r)=-2\, (n-1)\(1+(n+1)\,\sigma\)
\frac{\psi'}{\psi}.  \ee
The conditions for the auxiliary variables $W_\al$ used in the
Lindblom {\em et al} drivers are found from (\ref{W_al}) to be
$W_t(0,r)=W_r(0,r)=0$.

%%%%%%%%%%%%%%%%%%%%%%%%%%%%%%%%%%%%%%%%%%%%%%%%%%%%%%%%%%%%%%%%%%%%%
\section{Numerical Approach}
\label{sec_evolution}
%%%%%%%%%%%%%%%%%%%%%%%%%%%%%%%%%%%%%%%%%%%%%%%%%%%%%%%%%%%%%%%%%%%%
Here we describe our strategy for the numerical solution of 
the GH system (with a scalar matter source) in spherical symmetry.

%%%%%%%%%%%%%%%%%%%%%%%%%%%%%%%%%
\subsection{The numerical grid and the algorithm}
\label{sec_numerics}
\begin{figure}[t!]
  \centering \noindent
  \includegraphics[width=13cm]{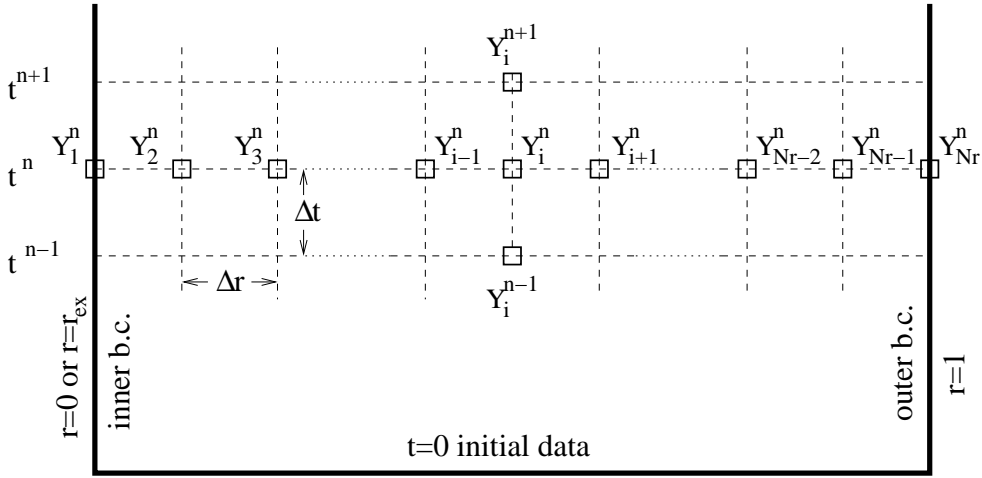}
  \caption[]{The compactified domain of integration, and the numerical
    lattice. Our finite difference scheme uses three levels in the
    time direction. }
  \label{fig_grid}
\end{figure}

%%%%%%%%%%%%%%%%%%%%%%%%%%%%%%%%%
We cover the $t$--$r$ plane by a discrete lattice denoted by
$(t^n,r_i)=(n\,\Delta t,i\,\Delta r)$, where $n$ and $i$ are integers
and $\Delta t$ and $\Delta r$ define the grid spacings in the temporal
and spatial directions, respectively.
We note that when we perform convergence studies, we keep the ratio
$\Delta t/\Delta r$ constant so that our numerical scheme is generally
characterized by a single discretization scale, $h$, which we can
conveniently identify with $\Delta r$.  As described in the next
section, the spatial domain is compactified, and hence a grid of
finite size $N_r$ extends from the origin to spatial infinity.  As
depicted in Fig.~\ref{fig_grid}, approximations to the dynamical
fields, collectively denoted here by $Y$, are evaluated at each grid
point, yielding the discrete unknowns $Y^n_i \equiv
Y(t^n,r_i)=Y(n\,\Delta t,i\,\Delta r)$.
In the interior of the domain, the GH equations and the gauge-driver
equations are almost always discretized using ${\cal O}(h^2)$ finite difference
approximations (FDAs), which replace continuous derivatives with the
discrete counterparts given in (\ref{FDA_t}) and (\ref{FDA_c}). As in
\cite{FP1,FP2} our scheme directly integrates the second-order-in-time
equations (i.e.~we do {\em not} rewrite the equations as a system
which is first order in time).
%%%%%%%%%%%%%%%%%%%%%%%%%%%%%%%%%%%%%%%%%%%%%%%%%%%%%%%%%%%%%%%%%%%%%%
%%%%%%%%%%%%%%%%%%%%%%%%%%%%%%%%%%%%%%%%%%%%%%%%%%%%%%%%%%%%%%%%%%%%%%

Following discretization, we thus obtain finite difference equations
at every mesh point for each dynamical variable.  Denoting any single
such equation as
\be
\label{FDA_eq}
\L_Y|^n_i=0.  \ee
we then iteratively solve the entire system of algebraic equations as
follows.

First, we note that for those variables that are governed by equations
of motion that are second order in time, our ${\cal O}(h^2)$
discretization of the equations of motion results in a three level
scheme which couples advanced-time unknowns at $t^{n+1}$ to known
values at retarded times $t^n$ and $t^{n-1}$.  In order to determine
the advanced-time values for such variables, we employ a point-wise
Newton-Gauss-Seidel scheme: starting with a guess for $Y^{n+1}_i$
(typically, we take $Y^{n+1}_i=Y^n_i$) we update the unknown using
\be
\label{Gauss-Seidel} {Y^{n+1}_i} \rightarrow {Y^{n+1}_i} - {\R_Y|^n_i
  \over\J_Y|^n_i}. \ee
Here, $\R_Y $ is the residual of the finite-difference
equation~(\ref{FDA_eq}), evaluated using the current approximation to
$Y^{n+1}_i$, and the diagonal Jacobian element is defined by
\be
\label{jacobian}
\J_Y|^n_i \equiv \frac{\pa \L_Y|^n_i}{\pa Y^{n+1}_i}.  \ee
In the cases where we used gauge drivers that involve $B$ and $W_\al$,
we found that an iteration based on an implicit Euler discretization
scheme of the corresponding first order equations performed
well.\footnote{The advantage of the implicit Euler method is that it is
  unconditionally stable and easy to implement. Although it is only
  first-order accurate---which does impact the overall convergence of the
  scheme when the Lindblom {\em et al} drivers are used---we have 
  found it useful to achieve our chief current goal of constructing
  stable numerical implementations for our GH system.}
  Specifically, writing any such equation
schematically as $\dot Y = f_Y(Y,\pa Y ,\dots)$, we update using
\be
\label{fully_implicit}
Y^{n+1}_i \rightarrow Y^{n-1}_i +2\Delta t\, f_Y|^{n+1}_i .  \ee
We iterate (\ref{Gauss-Seidel}) and (\ref{fully_implicit}) over all
equations until the overall residual norm\footnote{defined, e.g.~as a
  sum of absolute values of the individual residuals of the equations,
  $\R=\sum_Y |\R_Y|$.} falls below some specified convergence
threshold.

In order to inhibit high-frequency\footnote{``High-frequency'' refers
  to modes having a wavelength of order of the mesh spacing, $h$.}
instabilities which often plague finite difference equations such as
ours, we add explicit numerical dissipation of the Kreiss-Oliger type
\cite{KO} to our scheme.  Following \cite{FP1}, at every grid point
and for each dynamical variable we make the replacement
\be
\label{KO_filter}
Y_i \to Y_i-\eps_{\rm KO}\,d_i \ee
at both the $t^{n-1}$ and $t^n$ time-levels before updating the
$t^{n+1}$ unknowns.
Here, $d_i$ is defined by \be
\label{eta_KO}
d_i \equiv \frac{1}{16}\(Y_{i-2}-4\,Y_{i-1} +6\,Y_{i} -4\,Y_{i+1} +
Y_{i+2}\). \ee
and $\eps_{\rm KO}$ is a positive parameter satisfying $0\le \eps_{\rm
  KO} \le 1$ that controls the amount of dissipation.  An extension of
the dissipation to the boundaries \cite{FP1}, as well as to the black
hole excision surface (see Sec.~\ref{sec_ah} ), was also tried, but
was not found to have any positive effect. In fact, using
dissipation at the outer boundary usually resulted in late-time
instabilities in the code.

%%%%%%%%%%%%%%%%%%%%%%%%%%%%%%%%%%%%%%%%%%%%%%%%%%%%%%%%%
\subsection{Coordinates and boundary conditions}
\label{sec_coords_bc}
%%%%%%%%%%%%%%%%%%%%%%%%%%%%%%%%%%%%%%%%%%%%%%%%%%%%%%%%%

While the physical, asymptotically flat spacetime extends to spatial
infinity, in a numerical code one can only use grids of finite size.
A standard strategy to deal with this issue involves truncating the
solution domain by introducing an outer boundary at some finite radius
where approximate boundary conditions are imposed.  When such an
approach is adopted, it is then important to ensure that the computed
solutions do not depend sensitively on the truncation radius.
However, another technique which has been successfully used in
previous work in numerical relativity, see e.g.~\cite{FP2,Choptuik_BS},  involves
compactification of the spatial domain.  Paralleling the experience of
these earlier studies, we have found that compactifying the radial
direction and imposing the (exact) Dirichlet conditions
(\ref{asymp_bc}) at the edge of the domain works well, provided that
we use sufficient dissipation.  In particular, it is known that due to
the loss of resolution near the compactified outer boundary (assuming
a fixed mesh spacing in the compactified coordinate), outgoing waves
generated by the dynamics in the interior will be partially reflected
as they propagate towards the edge of the computational domain, and
these reflections will then to tend to corrupt the interior solution.
By adding sufficient dissipation one can damp the waves in the outer
region, attenuating any unphysical influx of radiation, and thus
enabling a meaningful use of compactification.

For the general case where we have more than one spatial dimension,
$X^i$, requiring compactification, we consider a transformation that
maps $X^i \in [0, \infty)$ onto $x^i \in [0,1]$,
\be \label{compact_coords} X^i = \zeta_i (x^i), \ee
where the $\zeta_i$ are monotonic functions, such that
$\zeta_i'(0)=1$, and which will have essential singularities at
$x^i=1$.
The field equations (\ref{EqHgab}-\ref{EqHPhi}) are discretized in the
compactified coordinates after we analytically remove the Jacobian of
the transformation (\ref{compact_coords}) in all the differential
operators.  The general replacement rule for first and second spatial
derivatives is $\pa_X =e_1 \pa_x$ and $\pa_{X}^2 = e_1^2 \pa_x^2 + e_2
\pa_x$, where $e_1\equiv 1/ \zeta'$ and
$e_2\equiv-\zeta''/(\zeta')^3$, so, for example, a typical term in
(\ref{EqHgab}-\ref{EqHPhi}), $\pa g_{t i}/\pa X^j$, would be replaced
with $(\zeta_j')^{-1} \pa g_{t i}/\pa x^j$.

In the spherically-symmetric calculations considered in this paper we
use a specific compactification
\be
\label{comact_r}
\tilde{r}=\frac{r}{1+r}, \ee
where the compactified $\tilde{r}$ ranges from $0$ to $1$ for values
of the original radial coordinate $r \in [0,\infty)$.  The boundary
conditions at $\tilde{r}=1$ are then imposed exactly: $g_{tt} =
-1,g_{tr}= 0, g_{rr}= 1, \lam=S = 0$, and $\phi= 0$. For the gauge
source functions we set $H_\al=0$,
as well as $W_\al = B =0$.  

We have previously described the boundary
(regularity) conditions at $\tilde{r}=r=0$ in 
Sec.~\ref{sec_axis_bc}.  Denoting by $Y^{n+1}_1$ the advanced-time value
at the origin for any of the variables, $\gtt, \grr$ and $H_t$ that
have vanishing derivative at $r=0$, we use the update
$Y^{n+1}_1=(4\,Y^{n+1}_2-Y^{n+1}_3)/3$, which is based on an $\O(h^2)$
backwards difference approximation (see (\ref{FDA_b})) of $\partial_r Y=\partial_{\tilde
  r}Y=0$.  For the quantities $\gtr$ and $H_r$, which are odd in $r$
as $r\to0$, we simply use $Y^{n+1}_1=0$.

As discussed in Sec.~\ref{sec_axis_bc}, we considered the
introduction of a new variable, $\lam$ (\ref{lambda_var}), to expedite
implementation of the regularity conditions involving $g_{rr}$ and
$S$.  However, in the calculations described below we have adopted a
simple method that does not involve $\lam$ and that works well in
spherical symmetry.
\footnote{However, we have checked that the scheme that uses $\lam$ performs 
remarkably well in our $2+1$ numerical implementation \cite{2Dcode} that
 generalizes the present $1+1$ work.}   
In this approach, we retain the original
variables $S$ and $g_{rr}$, and impose $g_{rr}' =0$ and $S =(1/2)
\log(g_{rr})$ at the origin. Then instead of determining $S^{n+1}_2$
(i.e.~the advanced value of $S$ at the next-to-extremal grid point)
from the corresponding discrete evolution equation, we perform the
update using the $\O(h^2)$ backwards FDA to the regularity condition,
$S'(t,0) =0$, namely $S^{n+1}_2=(3\, S^{n+1}_1 +S^{n+1}_3)/4$.

We must also maintain regularity at the origin for the auxiliary
functions $W_\al$ and $B$ that are used with some of the gauge driver
conditions.  We expand the metric functions in analytic Taylor series
around $r=0$ and substitute the expansions into the equations
(\ref{W_al},\ref{eqB_SS}) to arrive at
\bea
\label{W-B_axisbc}
\dot B &+& \eta_2\, B=0, \non \dot W_t &+&\gtt \(\eta\,
W_t\,\grr-(n+1)H_t''\)=0, \non \dot W_r &+& \gtt\(\eta\,
W_r\,\grr-H_r''\)=0, \eea
which we use to advance $B(t,0)$ and $W_\al(t,0)$ forward in time.
Operationally, the time-derivatives in the equations are replaced with 
the FDA expressions (\ref{FDA_t}) evaluated at $t^{n}$, and the spatial derivatives are replaced with one-sided 
versions (\ref{FDA_b}) evaluated at   $t^{n+1}$. The values of the functions $B(t^{n+1},0)$ and $W_\al(t^{n+1},0)$ 
are then algebraically found. 
%%%%%%%%%%%%%%%%%%%%%%%%%%%%%%%%%%%%%%%%%%%%%%%%%%%%%%%%%%%%%%%%%%%%%%%%%%
\subsection{Apparent horizon and excision}
\label{sec_ah}
As is well known from many theoretical studies (both closed-form and
numerical), a gravitational collapse process that concentrates
sufficient mass-energy within a small enough volume can lead to the
formation of a black hole.  In numerical calculations based on a
space-plus-time split, black hole formation is often inferred by the
appearance of apparent horizons.  We recall that an apparent
horizon is defined as the outermost marginally trapped surface, and
that a marginally trapped surface is one on which future-directed null
geodesics have zero divergence.  Specifically, given a surface with
outward-pointing spacelike unit normal, $s^\al$, embedded in a
hypersurface with future-directed timelike unit normal, $n^\al$, the
vanishing of the divergence, $\theta$, of the outgoing null rays
defined by $l^\al = s^\al +n^\al $ can be expressed as
\be \label{zero-divergence} \theta=(\gm^{\al\bt}-s^\al s^\bt)
\nabla_\al l_\bt = 0. \ee
In spherical symmetry we have $s^\al=g_{rr}^{-1/2} \, \pa_r$, and the
above equation can be written as\footnote{An
  alternative way to derive this result relies on the fact that the
  apparent horizon in spherical symmetry can be defined as a null
  surface located at constant radius.  Equating the
  time-derivative of the areal radius along null rays to zero, ${d (r\,e^{S})
    /d t}|_{l^\al} = r \,e^{S}\pa_t S +r\,e^S\,(1/r+\pa_r S)
  \(-g_{tr}/g_{rr}+\sqrt{g_{tr}^2/g_{rr}^2-g_{tt}/g_{rr}}\)=0$, where
  the expression in the second brackets is $dr/d t|_{l^\al}$, we recover
  the result in (\ref{zero-divergence-1d}). }
\be \label{zero-divergence-1d} \theta= r\, \pa_t S +(1+r\,\pa_r
S)\,\(-{g_{tr}\over g_{rr}}+\sqrt{{g_{tr}^2 \over
    g_{rr}^2}-{g_{tt}\over g_{rr}}}\) = 0, \ee
In numerical calculations, one can thus easily locate an apparent
horizon by simply searching for zeros of $\theta$: the position of the
outermost such zero then coincides with the location, $r_{\rm AH}$, of
the apparent horizon.

In our code we use excision to (dynamically) exclude from the
computational domain a region interior to the apparent horizon that
would eventually contain the black hole singularity.  The success of
this approach hinges on the observation that in spacetimes that
satisfy the null energy condition (such as those that we construct)
and assuming cosmic censorship, the apparent horizon is contained
within the event horizon, which ensures that the excluded region is
causally disconnected from the non-excised portion of the domain (see
\cite{thornburg_excision} and the references therein for further
discussion).  Operationally, once an apparent horizon is found, we
introduce an excision radius, $r_{\rm EX}$, that satisfies $r_{\rm
  EX}<r_{\rm AH}$, and such that all radial characteristics at
$r=r_{\rm EX}$ are pointing inwards. (We typically find 
$r_{\rm EX} \approx 0.4\,r_{\rm AH}$, where we again emphasize that 
$r$ is the coordinate radius.) 
This specific characteristic
structure eliminates the need for boundary conditions at $r_{\rm EX}$:
rather, advanced-time unknowns located on the excision surface are
computed using finite difference approximations to the interior
evolution equations, but where centered difference formulae are
replaced with the appropriate one-sided expressions given by (\ref{FDA_b}).
%%%%%%%%%%%%%%%%%%%%%%%%%%%%%%%%%%%%%%%%%%%%%%%%%%%%%%%%%%%%%%%%%%%%%%%%%%
\subsection{Spacetime diagnostics}
\label{sec_char}
%%%%%%%%%%%%%%%%%%%%%%%%%%%%%%%%%%%%%%%%%%%%%%%%%%%%%%%%%%%%%%%%%%%%%%%%%%
We employ several diagnostics in order to characterize the geometries
of the spacetimes we construct.

{\it Mass.}  Far away from an isolated system a natural radial
coordinate is defined by the asymptotic flatness of the 
spacetime, and the ADM mass of
the solution can be found from the asymptotic radial behavior of the
metric functions.  In spherical symmetry there is only one asymptotic
constant, $r_0$, that can be determined, for instance, from the fall-off of
$g_{tt}$: $g_{tt}\sim 1+r_0^{n-1}/r^{n-1}$. This constant is related to
the mass \cite{MP} by $M=n\,\Omega_n/(16\,\pi) r_0^{n-1}$, where
$\Omega_n=2 \pi^{((n+1)/2)}/\Gamma[(n+1)/2]$ is the surface area of a
unit $n$-sphere.

In addition, in spherical symmetry one can define a local mass
function, $m(t,r)$, sometimes called the {\em mass aspect}
\be \label{M_loc} m(r,t)= {n\,\Omega_n r^{n-1}\over 16\,\pi}
\(1-R_{,\al} R_{,\bt} g^{\al\bt} \), \ee
where $R=r\,e^S$ is the areal radius.
The mass aspect is negative inside a trapped (or anti-trapped) region, vanishes at its
boundaries and is positive outside in regular region.  It grows monotonically and
asymptotically coincides with the ADM mass.

\begin{figure}[t!]
  \centering \noindent
  \includegraphics[width=10cm]{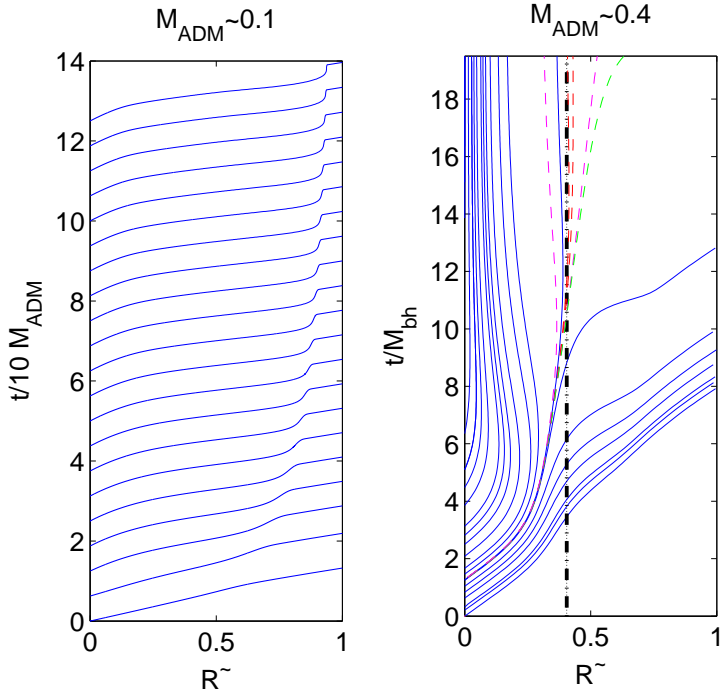}
  \caption[]{Outgoing null rays in the $t-\tilde{R}$ plane emitted
    from the origin at different times ($\tilde{R}$ is the
    compactified areal radius).  The left panel shows the geometry 
    generated by an
    initially origin-centered pulse of matter with $ \Phi_0=1.6$
    that disperses
    infinity. The presence of the matter deflects the outgoing null
    rays towards the origin, but the rays eventually escape to
    infinity. The motion of the pulse can clearly be traced.  The
    right panel shows the geometry generated by stronger initial data having
    $ \Phi_0=3.0$.  In this case the matter collapses to form a 
    black hole of mass,
    $M_{\rm BH}\simeq 0.3$: rays emitted before
    $t\simeq 1.25 M_{\rm BH}$ escape to infinity but the rays emitted
    after that time fall back to the origin. The null ray that separates the
    two regimes designates the event horizon and the thick dashed line is
    the asymptotic apparent horizon. The thin dashed lines are
    obtained by integrating (\ref{null_rays}) backward in time, and
    are attracted to the event horizon. }
  \label{fig_nullrays}
\end{figure}

{ \it Null geodesics.}  A convenient way to visualize the causal
structure of a spherically symmetric spacetime is to plot a family of
outgoing null rays, $l_\al$.  When plotted in the $t$--$R$ plane, the
slope, $dR/dt|_{l^\al}$, of an outgoing null geodesic is positive
outside the apparent horizon, and asymptotes to the flat-space value
of unity for large values of $R$.  Additionally, the slope vanishes at
the apparent horizon, concomitant with the vanishing of the outgoing
null divergence, and becomes negative inside the horizon.  All of
these features can be seen in Fig.~\ref{fig_nullrays}, where the
displayed lines are integral curves, ${\tilde R}(t;t_0)$. Here
${\tilde R}$ is the compactified areal radius, and the corresponding
uncompactified trajectory, $R(t;t_0)$, is defined by
\be
\label{null_rays}
R(t;t_0)=\int^t_{t_0}\[ \(-{g_{tr}\over g_{rr}}+\sqrt{{g_{tr}^2 \over
    g_{rr}^2}-{g_{tt}\over g_{rr}}}\)\frac{\pa R}{\pa r} + \frac{\pa
  R}{\pa t} \]\,dt'. \ee
Each curve thus represents the path of an outgoing null ray that is
emitted from the origin at a specific time, $t=t_0$.
%%%%%%%%%%%%%%%%%%%%%%%%%%%%%%
%%%%%%%%%%%%%%%%%%%%%%%%%%%%%%

{\it Event horizon.} In contrast to the local definition
(\ref{zero-divergence-1d}) of the apparent horizon, the event horizon
is a global concept: it is defined by outgoing null rays that neither
escape to future null infinity, nor fall into the black-hole
singularity.  Clearly, this definition requires knowledge of the
complete time evolution of the system, and hence, assuming a
calculation that is carried out for a finite amount of coordinate (or
proper) time, one cannot even in principle locate event horizons in
numerically-generated spacetimes. However, when a spacetime approaches
a stationary state, an approximate event horizon {\em can} be found.
We employ the method of Libson {\em et al} \cite{Libson} which is
based on the observation that if one integrates the geodesic equation
(\ref{null_rays}) backward in time, the event horizon becomes an
attractor for geodesics that either escape to future null infinity or
fall into the singularity at arbitrarily late times.  We have found
that in our simulations the event horizon is traced fairly well by the
time development of the apparent horizon. Again this can be seen in
Fig.~\ref{fig_nullrays}, where the thin dashed lines show the
trajectories obtained by integrating (\ref{null_rays}) backwards in
time, and starting with several initial radii.
%%%%%%%%%%%%%%%%%%%%%%%%%%%%%%%%%%%%%%%%%%%%%%%%%
\section{Results}
\label{sec_results}
%%%%%%%%%%%%%%%%%%%%%%%%%%%%%%%%%%%%%%%%%%%%%%%%
For concreteness, we restrict our numerical experiments to the case of
four-dimensional spacetimes, and take our matter source to be a
real, massless scalar field.  All of the results discussed here were
generated using an initial scalar field profile of the Gaussian
form~(\ref{Phi0}), with fixed values $r_0=0$ and $\Delta=0.6$, so that
the scalar pulse is always initially centered at the origin.  The
overall amplitude, $\Phi_0$, of the profile~(\ref{Phi0}) is then used
as a control parameter: variations of $\Phi_0$ produce varying
``strengths'' of initial data, and varying degrees of non-linearity in
the ensuing evolution.  In practice, the maximum value of
$2m(t,r)/R(t,r)$ (where $R$ is the uncompactified areal radius) that
is achieved in a given calculation is a useful indication of how
strong-field the evolution becomes.

We use the above notion of initial data strength to loosely define
three classes of solutions---within a given class we observe that the
overall dynamics of each of the scalar and gravitational fields are similar.
Specifically, we consider the following cases: (i) weak data, defined
by $\Phi_0 \lesssim 0.5$, yielding $\max_{t,r} 2 m/R\simeq 0.08 $;
(ii) intermediate data, having $0.5 \lesssim \Phi_0 \lesssim 1.6$, and
$\max_{t,r} 2 m/R\simeq 0.25 $, and (iii) strong data, with $\Phi_0
\gtrsim 1.6$ and $\max_{t,r} 2 m/R > 0.25 $.  While the first two
cases describe weakly and mildly gravitating scalar pulses,
respectively, which completely disperse in all instances, the strong
data generates spacetimes in which black holes form, or almost form
(i.e.~near-critical evolution, see~(\cite{Choptuik_critcollapse})).

We have also found it useful to use the total ADM mass, $M_{\rm ADM}$,
of the spacetime---which can be computed at $t=0$---to normalize
certain numerical parameters.  In particular, we set
the parameters of the gauge driver (\ref{FP_gauge},\ref{FP_gauge_m})
using $\xi_1={\xi_{10}}/M_{\rm ADM}^2$, $\xi_2={\xi_{20}}/M_{\rm ADM}$
and $\xi_3={\xi_{30}}/M_{\rm ADM}^2$, where the ``bare'' values,
$\kappa_0$, $\xi_{10}$, $\xi_{20}$ and $\xi_{30}$ are generally held
fixed as $\Phi_0$ is varied.  Moreover, and as discussed in more
detail below, we find that the accuracy of our results is improved if
the constraint damping term asymptotically vanishes at large spatial
distances. Accordingly, we typically multiplied $\kappa$ by the factor $2M_{\rm
  ADM}/R$.

Because we use, at least in large part, a time-explicit finite
difference scheme, we expect restrictions on the ratio $\lambda_C
\equiv \Delta t/\Delta r$ (the Courant factor) that can be used while
maintaining numerical stability.  For the case of harmonic gauge, we
found that values of $\lambda_C$ satisfying $0.01 \lesssim \lambda_C
\lesssim 0.8$ generated stable solutions with roughly constant
accuracy, although somewhat stronger numerical dissipation was
required to stabilize runs that used larger values of $\lambda_C$ in
that interval.  In the results discussed below we have typically taken
$0.3\lesssim \lambda_C \lesssim 0.6$ for weak and intermediate data,
and $0.1\lesssim\lambda_C\lesssim 0.2$ for the evolution of strong
data.
We further found that when any of the other gauge drivers were
adopted, smaller Courant factors (relative to the harmonic case) were
required. In those cases our results  were generally computed using
$0.05\lesssim\lambda_C\lesssim 0.1$.
Typically, in cases where $\lambda_C$ was taken too
large, we observed amplification and dominance of numerical errors near the
origin: this lead to high frequency oscillations and, eventually, to
divergence of the numerical solution.

Another crucial numerical parameter is the Kreiss-Oliver dissipation
factor, $\eps_{\rm KO}$, which we generally set according to $0.1
\lesssim \eps_{\rm KO} \lesssim 0.7$.  Finally, it is important to
note that we found that optimal values of both $\lambda_C$ and
$\eps_{\rm KO}$ were dependent on the spatial resolution:
specifically, as $\Delta r\to 0$ somewhat smaller values of
$\lambda_C$, as well as larger values of $\eps_{\rm KO}$ were usually
required.
The lowest and highest resolution runs reported in this paper typically had
$\Delta r = 1/64$ and $\Delta r = 1/8192$, respectively: runs with
$\Delta r = 1/8192$ generally required $\lambda_C=0.05$ and 
$\eps_{KO}=0.7$ for stability.

Many of the coordinate conditions discussed and employed in this paper
are characterized by several adjustable parameters, and we have by no
means carried out exhaustive parameter space surveys in all cases in
an attempt to optimize parameter settings.  Rather, our more limited
numerical experimentation indicates that with a certain amount of
tuning of the parameters, it does seem possible, at least in
principle, to simulate various interesting situations.  Our intent
here is chiefly to document the overall behavior of several gauge
conditions as well as to explore some of the effects that specific
parameters of the gauge drivers have on the evolution.  Given this
primary goal, we also defer most of our discussion of code convergence
and accuracy to Sec.~\ref{sec_convergence}.

%%%%%%%%%%%%%%%%%%%%%%%%%%%%%%%%%%%%%%%%%%%%%%%%
\subsection{Weak data}
\label{sec_weakdata}
%%%%%%%%%%%%%%%%%%%%%%%%%%%%%%%%%%%%%%%%%%%%%%%%
In this section we consider the evolution of weak initial data for
which $\Phi_0 \lesssim 0.5$, yielding $M_{\rm ADM} \lesssim 0.01$ and
$\max_{t,r} 2 m/R\lesssim 0.08$.
In this case there is little interaction between the scalar and
gravitational fields, the scalar pulse entirely disperses to infinity,
and we find that essentially any of the gauge conditions described
above can be used to produce long-term stable evolution.  For this
type of data we use $\eps_{\rm KO}\simeq 0.1$ for the Kreiss-Oliger
dissipation parameter, finding that larger values have detrimental
consequences for stability.
However, even with dissipation and constraint damping,
we find that numerical errors 
eventually do grow---on a time scale of order $t>10^4 M_{\rm ADM}$---and
cause the code to crash.

We find that the effect of the constraint damping term depends on
whether $\kappa$ is fixed or allowed to vary over the integration
domain. For fixed $\kappa$, it is essential to take $\kappa_0 > 0.01$,
otherwise high-frequency oscillations quickly ruin convergence.
However, if the damping is too strong, instabilities are also
triggered.  In fact, we find that the optimal damping parameter is
related to the typical scale over which the scalar field varies.  For
the Gaussian initial data that we consider, this scale is $\Delta$, so
we take $\kappa \simeq \Delta^{-1}$.  (This observation holds for
intermediate strength data as well, as can be seen in 
Fig.~\ref{fig_kappa2}.)  On the other hand, when we take
$\kappa=\kappa(r)$, and specifically for the choice $\kappa=\kappa_0
(2 M_{\rm ADM} /R)$ mentioned previously, we find that the results are
relatively insensitive to the value of $\kappa_0$, provided $\kappa_0
\lesssim 100 \Delta^{-1}$.  For larger values of $\kappa_0$
instability is again usually observed.

Our experiments with the gauge drivers proposed by Lindblom {\em et
  al}, have focused on the specific Bona-Masso slicing condition for
which $f(\al)= 2/\al$, corresponding to $1+\log$ slicing.  However,
for weak data, we find that other choices of $f$
(such as $f(\al)=2\al, \al^2$ and $10/\al^2$, to list a few that we have
tried) produce qualitatively similar results.

Considering the conditions that determine the shift, we find that the
$\Gamma$-driver condition performs somewhat better than
$\Gamma$-freezing, with the former allowing the evolution to be
controlled for a longer amount of time.
There was only mild dependence on the gauge-driver parameters,
$\mu_{1,2}, \eta_{1,2}, q_s,g_n, \sigma $ and $\nu$, provided they
are all taken in the range $0.01$--$10$ in units of $M_{\rm
  ADM}$.

In order to assess the performance of the coordinate conditions in
driving the source functions to the target functions, we first follow
\cite{Lindblometal2} and define the weighted $L_2$-norm, $|Y|$, of a
function $Y$ as follows\footnote{The integrals are evaluated on our
  fixed mesh using the trapezoidal rule. },
\be
\label{Lind_norm}
|Y| = \(\frac{\int e^{2S} Y^2 {r}^2 \sqrt{\grr} d{r}}{\int e^{2S}
  {r}^2 \sqrt{\grr} d{r}} \)^{1/2}.  \ee
A similar, if somewhat less smooth norm, which we also use here, can be
defined as
\be
\label{l2norm}
|Y|_{L_2}= \frac{1}{N_r} \sqrt{\sum_{i=1}^{N_r} Y^2}.\ee

\begin{figure}[t!]
  \centering \noindent
  \includegraphics[width=12cm]{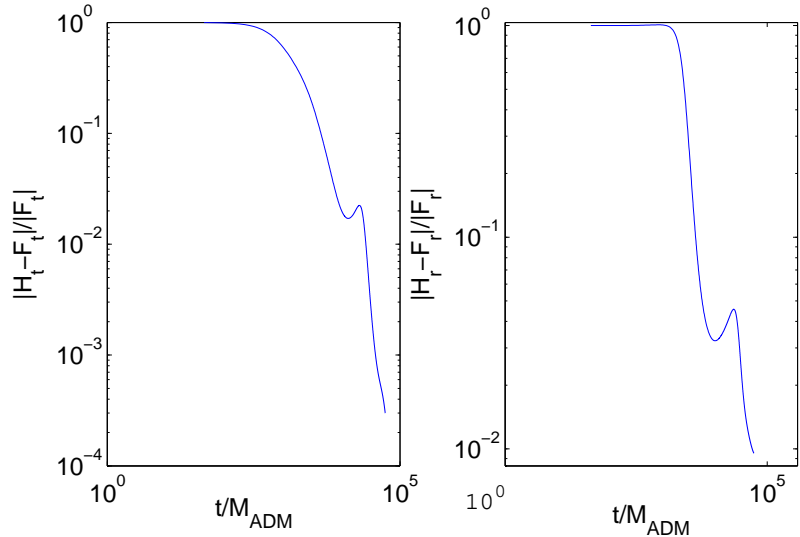}\vspace{0.3cm}
  \caption[]{The behavior of gauge drivers for the case of Bona-Masso
    slicing with $f(\al)=2/\al$ (left) and the $\Gamma$-driver shift
    condition (right) in the weak field regime, $\Phi_0\simeq 0.1$ }
  \label{fig_Fs_weakdata}
\end{figure}

Fig.~\ref{fig_Fs_weakdata} shows the weighted norms of the
differences between the actual and target source functions from a
typical weak-field simulation.  It is evident from these plots that
the drivers successfully drive the source functions $H_\al$ towards
the target functions $F_\al$ as the evolution proceeds.

We now continue to discussions of the evolution of intermediate- and
strong-field data, where the results are more sensitive to the
specific driver used, as well as to the parameter settings for any given
driver.
%%%%%%%%%%%%%%%%%%%%%%%%%%%%%%
%%%%%%%%%%%%%%%%%%%%%%%%%%%%%%

%%%%%%%%%%%%%%%%%%%%%%%%%%%%%%%%%%%%%%%%%%%%%%%%
\subsection{Intermediate data}
\label{sec_interdata}
%%%%%%%%%%%%%%%%%%%%%%%%%%%%%%%%%%%%%%%%%%%%%%%%
Here we consider evolutions characterized by $0.5 \lesssim \Phi_0
\lesssim 1.6$, where $M_{\rm ADM} \lesssim 0.1$ and $\max_{t,r} 2
m/R\lesssim 0.25$. First, for this strength of data, we have found that the pure harmonic
and GH gauges (\ref{FP_gauge_SS}-\ref{FP_gauge_SS_m}) perform
comparably.
With both choices, we are typically able to accurately trace the
evolution of the initial data for times of the order of 100--600
$M_{\rm ADM}$, with increasing
resolution resulting in increased maximum evolution time.

The causal structure of the spacetime from a typical intermediate
strength computation is displayed in the left panel of
Fig.~\ref{fig_nullrays}.  We recall
that in this figure the curves represent trajectories of outgoing null rays that are
emitted at regular intervals (in coordinate time) from $r=0$.  As the
evolution proceeds, the pulse, which is initially centered at the
origin, disperses to infinity. The outgoing null rays are bent towards
the origin by the presence of the matter and asymptotically become
straight lines with unit slope in the $r-t$ plane.  The position of
the scattered pulse of scalar field can be traced through the location
of the ``ripple'' in each curve, i.e.~at the positions where the
outgoing null geodesics suffer the most deflection.

We will discuss issues of code convergence
and accuracy in more detail in Sec.~\ref{sec_convergence}.
However, we note here that constraint norms, $|M_\al|_{L_2}$,
defined by~(\ref{Mal_4D}) and
computed, for example, using either (\ref{Lind_norm})
or (\ref{l2norm}) provide a basic indication of the accuracy of our
numerical method.
For the calculation depicted in Fig.~\ref{fig_nullrays} that uses
a medium resolution, $\Delta r=1/1024$, we find the initial norms
$|M_\al|_{L_2} $of order $10^{-4}$, which for roughly the first half
of the evolution then decrease to values of $10^{-5}$--$10^{-6}$.
Thereafter we observe a slow increase in the size of the constraints
although---except for the last few time steps before the code fails---
$|M_\al|_{L_2} $ remain well below the $10^{-3}$ level.  Moreover,
we generally observe the expected quadratic convergence of
$|M_\al|_{L_2} $ as the finite difference mesh is refined.

\begin{figure}[t!]
  \centering \noindent
  \includegraphics[width=11cm]{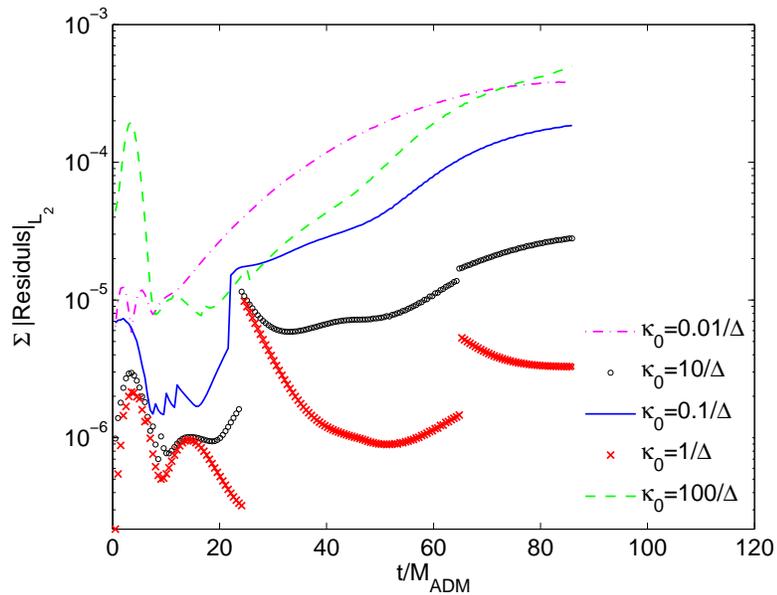}
  \caption[]{ Illustration that the characteristic behavior of the $L_2$ 
    residuals of
    the evolution equations depends on the value of the damping parameter. 
    Excessive or insufficient damping degrades convergence, or leads to
    divergence.  The optimal range for the damping parameter is
    $\kappa_0 \sim \O(1)/\Delta$, where $\Delta$ is the typical length
    scale in the problem.  }
  \label{fig_kappa2}
\end{figure}

Another basic indication of numerical accuracy is provided by the the
sum of the norms defined by (\ref{l2norm}) of the residuals of the dynamical
equations, $|\R|_{L_1}=\sum_Y |\R_Y|$, where $\R_Y$ is the FDA
residual of the equation that governs the field $Y$.  
Fig.~\ref{fig_kappa2} shows the behavior of $|\R|_{L_2}$ as a function of
the damping parameter, $\kappa_0$, for calculations with $N_r=513$
(moderate resolution), $\Phi_0=1.6$, and where $\kappa=\kappa_0 (2
M_{\rm ADM}/R)$.  As already noted in the discussion of the weak field
results, the sizes of the constraint and equation residuals tend to
be minimized when $\kappa_0$ is comparable to the inverse of the
typical length scale of the problem, i.e.~to $\Delta^{-1}$ for our
initially Gaussian data.  This is apparent in the figure, which shows
that for $\kappa_0 =0.5/\Delta$, the residuals remain on the order of
$10^{-5}$.
%%%%%%%%%%%%%%%%%%%%%%%%%%%%%%
%%%%%%%%%%%%%%%%%%%%%%%%%%%%%%

We next experiment with the Lindblom {\em et al} drivers, and find
that while for $\Phi_0 < 0.7$ the dynamics of $H_\al$ and $F_\al$ is
qualitatively similar to that in the weak field regime (shown in
Fig.~\ref{fig_Fs_weakdata}) and essentially independent of the
parameters of the gauge drivers, for $\Phi_0>0.7$ the convergence of
the source functions, $H_\al$, towards the target sources, $F_\al$,
has stronger dependence on the parameter settings.  The most pronounced
feature in this regime is that the drivers succeed in forcing
$H_\al\to F_\al$ only on the length-scale set by the parameter
$\mu_1$.  In particular, when we start with initial data that has
$H_\al=F_\al$, we find that for large values of $\mu_1$ the source
functions remain close to their targets for a a few tens of $M_{\rm
  ADM}$, after which high-frequency oscillations destroy the matching.
Conversely, starting from the same initial set up, but taking $\mu_1$
very small, we observe that the source functions quickly deviate from
the targets and never approach them in the subsequent evolution.

\begin{figure}[t!]
  \centering \noindent
  \includegraphics[width=14cm]{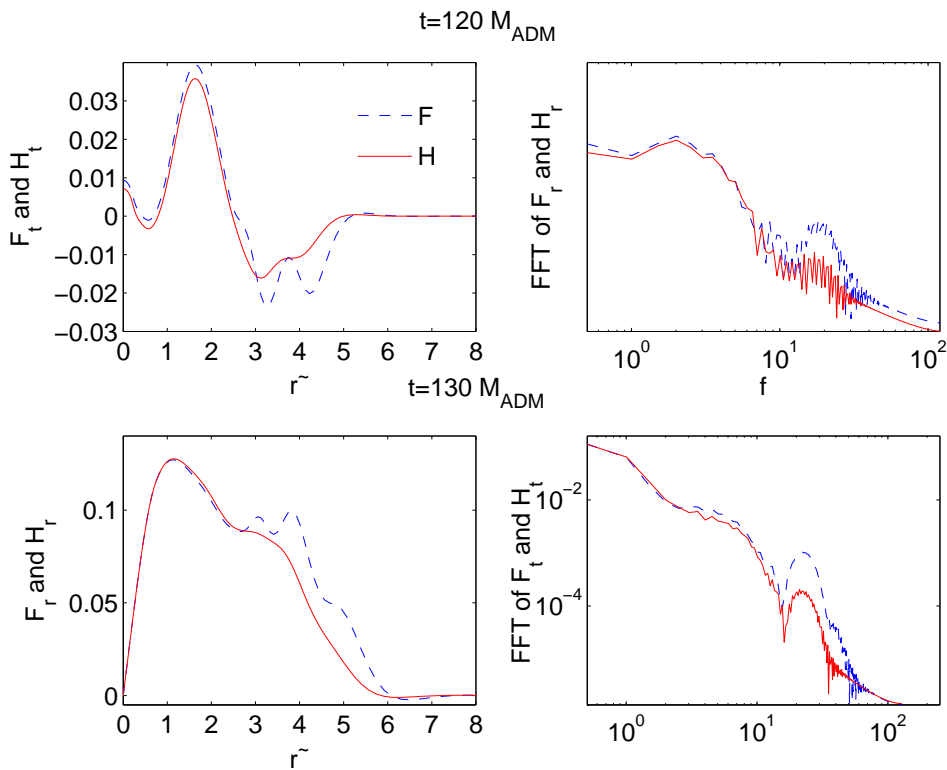}
  \caption[]{The source functions $H^\al$, the target functions
    $F^\al$, and their Fourier transforms at two instants. We begin
    with initial data satisfying $H_\al=F_\al$.  
    Within a few dynamical times the
    functions deviate, but subsequently are driven towards each other.
    After a time of $30-50 M_{\rm ADM}$ they match
    on length-scales of order $1/\mu_1$. This is illustrated by
    the spatial spectral decomposition shown in the right panels: while the
    lower frequencies of the functions match closely, the
    higher-frequency components do not. }
  \label{fig_Fs_intermdata}
\end{figure}

Given this observation, and given that our Gaussian initial data
generates an evolution characterized by a length scale,
$\Delta$, it is thus reasonable to take $\mu_1\simeq 1/\Delta$ in an
attempt to enforce the desired gauge conditions on that scale.
Results from such a computation are shown in Fig.~\ref{fig_Fs_intermdata}, 
which displays the source and target
functions, as well as their Fourier transforms, from the evolution of
initial data with $\Phi_0=0.9$.  The calculations were performed using
target slicing of the Bona-Masso type with $f(\al)=2/\al$, and target
$\Gamma$-driver shift conditions with $\mu_1=1.3$ (recall that
$\Delta=0.6$ for all of the computations described here).  In
addition, here, and for all of the results discussed in this section,
we used $\mu_2=\eta=\eta_2=1 $, $q_n= q_s=0.5,\sigma=-1/3$ and
$\nu=0.7$.  In contrast to the case of $\mu_1$, we find that the
calculations are not too sensitive to the settings of these
parameters, so long as their values are all of order unity.
In this simulation we begin
    with initial data satisfying $H_\al=F_\al$.  
    Within a few dynamical times the
    functions deviate, but as  Fig.~\ref{fig_Fs_intermdata} demonstrates 
the functions are subsequently driven towards each other, when 
the source functions start resembling the targets on the spatial scales
$1/\mu_1$. Notice that the high-frequency spatial variations of the
target $F^\al$'s are not replicated by the source functions.  Similar
behavior was originally observed in \cite{Lindblometal2} for
perturbations on a given background.
%%%%%%%%%%%%%%%%%%%%%%%%%%%%%%
%%%%%%%%%%%%%%%%%%%%%%%%%%%%%%

\begin{figure}[t!]
  \centering \noindent
  \includegraphics[width=12cm]{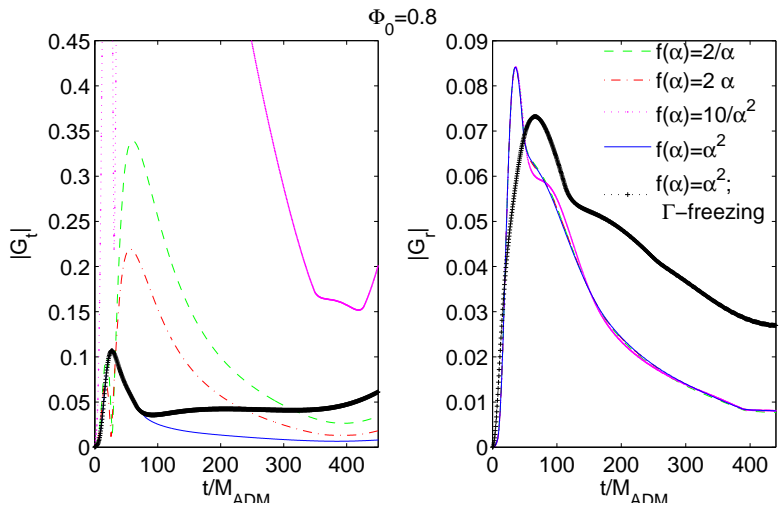}
  \caption[]{The norms of the functions $G_t$ and $G_r$, that must
    vanish when a specific gauge to which they correspond is
    approached. The norms decrease in time in a way that depends on
    the details of the gauges and the parameters of the drivers. In
    the simulations shown, we mostly use Bona-Masso slicing with
    various choices of $f(\al)$, and $\Gamma$-driver shift conditions,
    except for the data plotted with $+$-symbols that was obtained
    using a $\Gamma$-freezing shift condition.  In all cases we are
    able to drive the norms to a level of about $0.01$.}
  \label{fig_Gs_intermdata}
\end{figure}

The manner in which the coordinate conditions evolve in time for this
calculation is shown in Fig.~\ref{fig_Gs_intermdata}, which depicts
the norms of the functions $G_t$ and $G_r$, defined by~(\ref{Gn_K0}) or (\ref{Gn_BM}),
and (\ref{Gi_Gamma-freezing}) or (\ref{Gi-Gamma-driver-1}).
As described in Sec.~\ref{sec_gauge_conditions}, enforcing a
particular gauge is equivalent to driving these functions to zero.
Since we begin with initial conditions in which the gauge is exactly
fixed, the norms of $G_t$ and $G_r$ are initially zero.  Then on a
timescale of order several tens of $M_{\rm ADM}$, the norms grow to
some maximum value, after which they decrease slowly.  The details
depend on the particular coordinate choices, as well as on the
settings of the driver parameters, but usually it is possible to drive
the $L_2$-norms of $G_t$ and $G_r$ to the level of about $0.01$.
%%%%%%%%%%%%%%%%%%%%%%%%%%%%%%
%%%%%%%%%%%%%%%%%%%%%%%%%%%%%%

Although for smaller initial pulse amplitudes ($\Phi_0 \lesssim 1.0$)
we managed to find parameters for the Lindblom {\em et al} drivers
that asymptotically fix the desired gauges, we find that for
larger amplitudes the effectiveness of the drivers degrades, and for
$\Phi_0 \gtrsim 1.0 $ we could not find parameter settings that
enforce any of the specific gauges.
This does not necessarily mean that the code diverges: indeed, the
evolution often proceeds, but the behavior of the source function is
rather arbitrary.  In this regime we find that the evolution systems based
on the Lindblom {\em et al} drivers tend to be more dynamical and less
stable than one that uses simple drivers such as (\ref{FP_gauge_SS}).

%%%%%%%%%%%%%%%%%%%%%%%%%%%%%%%%%%%%%%%%%%%%%%%%
\subsection{Strong data and black hole formation}
\label{sec_strongdata}
%%%%%%%%%%%%%%%%%%%%%%%%%%%%%%%%%%%%%%%%%%%%%%%%
Increasing the initial amplitude, $\Phi_0$, of the scalar pulse leads
to increasingly strong curvature in the development of the initial
data.  As expected, above a critical value---in the current case,
$\Phi_0\sim 2.15$---black holes form, as signaled by the appearance of
apparent horizons.  We recall that we have already used the
trajectories of outgoing null geodesics to schematically display the
causal structure of a typical black hole geometry in the right panel
of Fig.~\ref{fig_nullrays}.

\begin{figure}[t!]
  \centering \noindent
  \includegraphics[width=10cm]{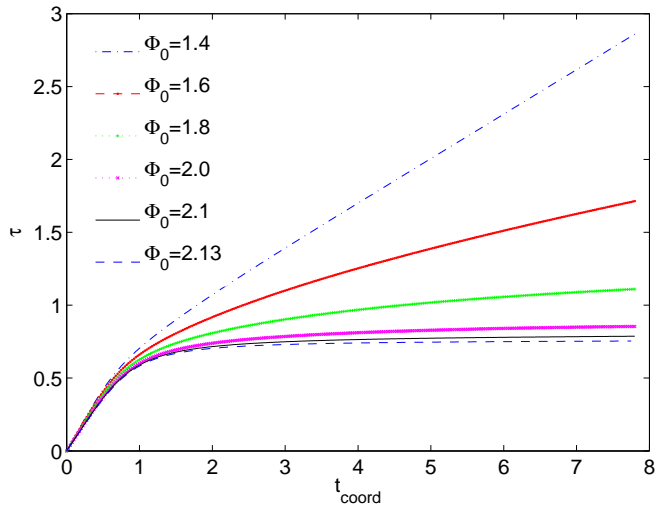}
  \caption[]{Proper time (\ref{tau}) at the origin in harmonic
    evolution as a function of coordinate time for several initial
    data strengths. The evolution slows down for stronger data and it
    effectively freezes for near critical data.}
  \label{fig_tau}
\end{figure}

Our first set of numerical experiments in the strong-field regime
compares subcritical evolution ($\Phi_0\lesssim 2.15$) in pure
harmonic coordinates to that in the generalized harmonic gauge given
by~(\ref{FP_gauge_SS}).  A generic feature of purely harmonic
evolution in this case is a fairly quick collapse of the lapse
function towards zero values near and at $r=0$.  As a result the
evolution in the central region (where the pulse is concentrated)
effectively freezes, and the scalar field remains present near $r=0$
even at late (coordinate) times.  This is demonstrated in
Fig.~\ref{fig_tau}, which shows the evolution of central proper time
\be
\label{tau}
\tau(t) \equiv \int^t_0 \al(t',0) dt', \ee
as a function of the strength of the initial data.
%%%%%%%%%%%%%%%%%%%%%%%%%%%%%%
%%%%%%%%%%%%%%%%%%%%%%%%%%%%%%

On the other hand, and in accordance with the previous experience of
Pretorius \cite{FP2}, we are able to use the generalized harmonic
gauge condition~(\ref{FP_gauge_SS})
to inhibit the collapsing of the lapse.  Specifically, we use $\al_0=1$ and $q=3$
in~(\ref{FP_gauge_SS}), and experiment with various values for $\xi_1$
and $\xi_2$.  In addition, motivated by an observation that we can
more stably evolve subcritical data by gradually ``turning-off'' the
gauge driving at late times, we actually replace $\xi_{1}$ and
$\xi_{2}$ in~(\ref{FP_gauge_SS}) by $(\xi_{10}/M_{ADM}^2)/(1+s\,t^p)$ and
$(\xi_{20}/M_{ADM})/(1+s\,t^p)$, respectively, where $p$ and $s$ are additional
positive parameters.  In practice, we have usually taken $p=1$,
leaving $s$ free to control the rate at which the gauge driving is
disengaged.

\begin{figure}[t!]
  \centering \noindent
  \includegraphics[width=12cm]{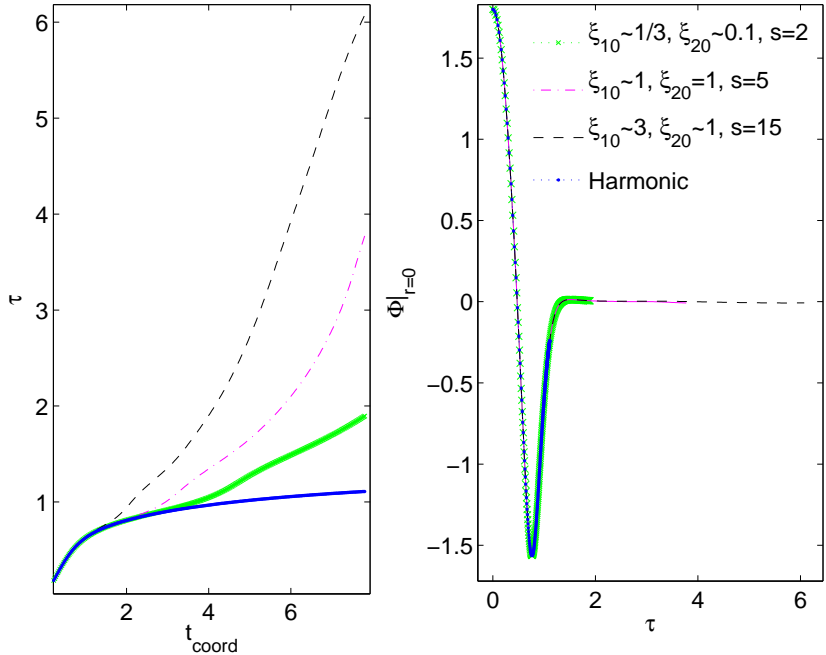}
  \vspace{0.1cm}
  \caption[]{Left panel: the proper time (\ref{tau}) at $r=0$ from
    an evolution that uses gauge conditions (\ref{FP_gauge_SS}) with
    $\Phi_0=1.8, H_r=0, q=3$, and $\xi_1$ and $\xi_2$ additionally
    divided by $1+s\,t$. Right panel: the amplitude of the scalar field at
    the origin from the same simulations.  While in harmonic gauge the
    evolution freezes near $r=0$, in the dynamical gauge
    (\ref{FP_gauge_SS}) it continues.}
  \label{fig_tau-xi}
\end{figure}

Results from calculations with $\Phi_0=1.8$ ($M_{\rm ADM} \simeq
0.125$) and using several sets of values for $\xi_{10}$, $\xi_{20}$
and $s$ are shown in Fig.~\ref{fig_tau-xi}.  The plots clearly show
how judicious choice of the parameters can prevent the collapse of the
lapse.  Through experiments with various subcritical initial data
sets we find that parameter values $1 \lesssim \xi_{10} \lesssim 5$
and $0.5 \lesssim \xi_{20} \lesssim 2$ produce good results.  However,
in order to keep the lapse from collapsing for initial data very close
to criticality, we generally needed to increase both $\xi_{10}$ and
$\xi_{20}$ by factors of up to 10, while simultaneously increasing $s$
(to values of order 50) and taking $p=2$ or $3$.
For instance, simulations that use $2049$ spatial grid points and
the driver (\ref{FP_gauge_SS}) with the parameters tuned to
$\xi_{10}=50$,$\xi_{20}=30$, $s=36 $ and $p=2$ allowed us to explore the
dynamics of solutions with $\Phi_0 =2.1465 \pm 0.0005$
without encountering a collapsing lapse. Unfortunately this is 
not close enough to the threshold amplitude for us to be 
able to observe in detail the
distinctive features of scaling and echoing known to appear
in the near-critical regime of this model \cite{Choptuik_critcollapse}.
%%%%%%%%%%%%%%%%%%%%%%%%%%%%%%
%%%%%%%%%%%%%%%%%%%%%%%%%%%%%%

We end our discussion of subcritical strong-field evolution with two
observations.  First, we note that while we have investigated the use
of dynamical conditions such as~(\ref{FP_gauge_SS_m}) for $H_r$, the
spatially harmonic choice, $H_r=0$, is simpler to implement, and
apparently more stable in this regime.  Secondly, although we have
experimented extensively with the Lindblom {\em et al} drivers in this
context, we have not been able to find parameter settings that prevent
coordinate pathologies (premature collapse of the lapse) from quickly
developing for near-critical evolutions.

We now turn to the case of supercritical evolutions, which are
characterized by the formation of black holes.  As described in
Sec.~\ref{sec_ah}, we have implemented black hole excision techniques
in our code: however, due to the strong singularity avoidance property
of pure harmonic gauge, as well as the generalized harmonic
modifications~(\ref{FP_gauge_SS}-\ref{FP_gauge_SS_m}), we can also
perform computations in which black holes form and are evolved for
some amount of time, but where excision is {\em not} used.

%%%%%%%%%%%%%%%%%%%%%%%%%%%%%%
\begin{figure}[t!]
  \centering \noindent
  \includegraphics[width=11cm]{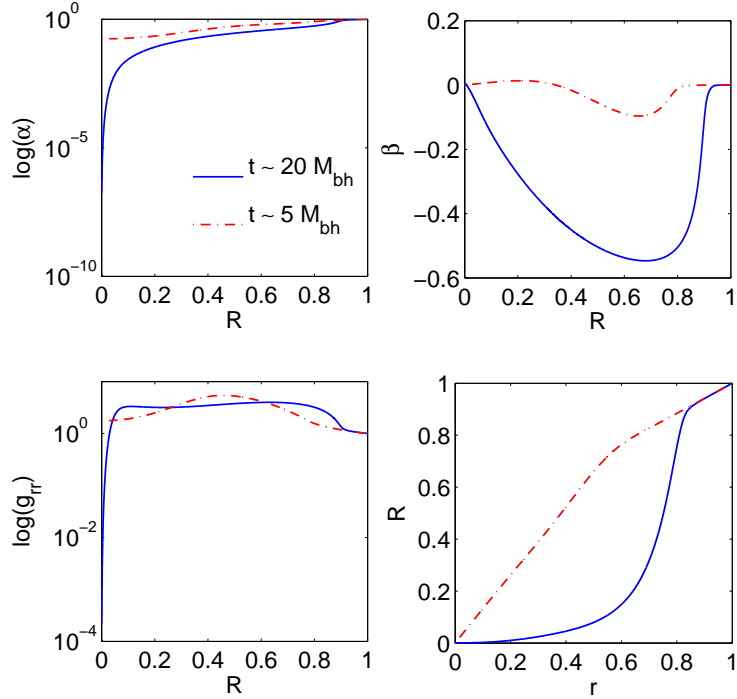}
  \caption[]{Illustration of the geometry of black hole formation 
    without excision.
    The metric functions remain regular all the way to the origin,
    where the functions tend to zero.  The black hole which forms
    has a mass
    $M_{\rm BH}\simeq 0.335$ with a horizon at ${\tilde R}\simeq
    0.4$ in the compactified areal radial coordinate. 
    Note that the lapse collapses
    at the origin, freezing the evolution there.  }
  \label{fig_noex}
\end{figure}
%%%%%%%%%%%%%%%%%%%%%%%%%%%%%%
%%%%%%%%%%%%%%%%%%%%%%%%%%%%%%
\begin{figure}[t!]
  \centering \noindent
  \includegraphics[width=9cm]{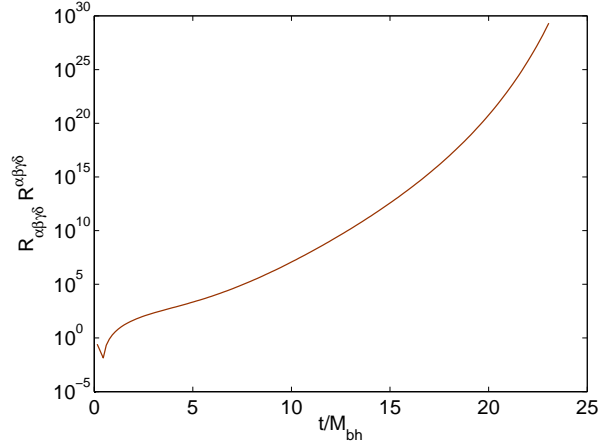}
  \caption[]{Time plot of the central value of the Kretschmann scalar,
    showing indefinite growth which signals the development of a 
    curvature singularity.  }
  \label{fig_noex_curv}
\end{figure}
%%%%%%%%%%%%%%%%%%%%%%%%%%%%%%
For example, Fig.~\ref{fig_noex} shows metric functions from a
calculation with $\Phi_0=3.0$ that uses pure harmonic gauge with no
excision.
We infer the formation of a black hole by the appearance of an
apparent horizon, which at the end of the simulation is located at a
compactified areal radius ${\tilde R}_{\rm AH}\simeq0.4$.  We can then
estimate the mass of the black hole at that time from the apparent
horizon location: $M_{\rm BH}= 0.5\, R_{\rm AH} \simeq 0.34$, and note
that the total ADM mass in this case is $M_{\rm ADM}\simeq 0.41$.  An
apparent horizon is first detected at $t\simeq 1.25 M_{\rm BH}$ and
Fig.~\ref{fig_noex} displays the metric functions at two instants:
$(i)$ $t\simeq 5 M_{\rm BH}$ (dashed lines), and $(ii)$ $t \simeq 20
M_{\rm BH}$, which is shortly before the simulation crashes (solid
lines).  For this specific calculation we used $4097$ spatial grid
points, and, at the time of the code crash, the values of the temporal
component of the metric, $g_{tt}$, near the origin are of order
$10^{-15}$ (corresponding to lapse values of order $\sim
10^{-7}$).  Despite the fact that all of the metric components
displayed in Fig.~\ref{fig_noex} are tending towards zero at the
origin at late times, the functions remain smooth and regular
throughout the evolution.  Fig.~\ref{fig_noex_curv} plots central
values for the Kretschmann scalar,
$R_{\al\bt\gm\delta}R^{\al\bt\gm\delta}$, as a function of time.  The
apparent divergence of this geometric quantity indicates the
development of a curvature singularity.

We have found that the use of excision can somewhat extend the
duration of our simulations of black hole spacetimes.  For comparison,
a run with the same parameters enumerated above, but employing
excision, lasted for as long as $\sim 60 M_{\rm BH}$.  We recall that
our simple approach to excision has been described in Sec.~\ref{sec_ah}, 
and note that in practice we have typically chosen the
excision radius, $r_{\rm EX}$, to satisfy $r_{\rm EX} \leq 0.4 r_{\rm
  AH}$.
The rest of the results described in this section were obtained in
simulations with excision.
%%%%%%%%%%%%%%%%%%%%%%%%%%%%%%
\begin{figure}[t!]
  \centering \noindent
  \includegraphics[width=12cm]{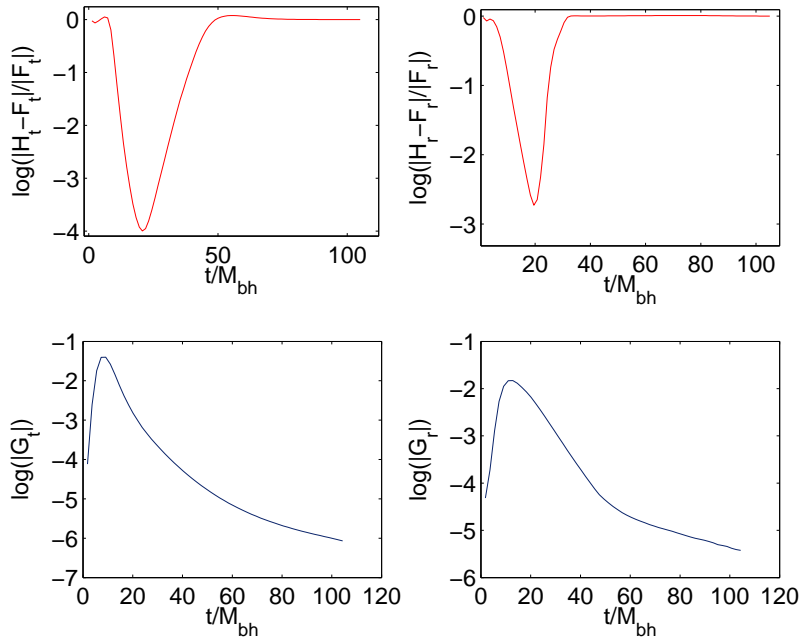}
  \caption[]{A strong data simulation, $\Phi_0=3$.  Shown are $L_2$-norms (\ref{Lind_norm}) of the normalized
    deviations between the target and source
    functions (top), and the gauge functions, $G^\alpha$, which vanish when 
    the desired gauge is achieved (bottom).  The source functions, $H_\al$,
    approach the targets, $F_\al$, uniformly soon after the horizon
    forms at $t\simeq 3\, M_{\rm BH}$ and follow them closely until low
    frequency variations in $F^\al$, induced by the matter outside the 
    black hole,
    develop and destroy the uniform match. The sources replicate the
    targets (not shown) on the scale defined by $\mu_1^{-1}$ which we
    take to be of order of $R_{\rm AH}$.  }
  \label{fig_Fs-Gs_Phi0=3}
\end{figure}
%%%%%%%%%%%%%%%%%%%%%%%%%%%%%%

Although we are able to avoid the central singularity using excision,
it is clear from our calculations that the harmonic coordinate system
continues to evolve in a highly non-trivial manner after excision is
initiated.  This dynamics in the coordinates causes, or is at least
associated with, two main problems.  First, the resulting coordinate
system does not approach a stationary state: in particular, the
coordinate position of the apparent horizon evolves with time.
Specifically, after formation, the horizon expands outwards and
consumes most of the numerical grid.  Eventually then, the portion of
the spacetime outside the horizon---which we recall extends to spatial
infinity due to our use of a compactified coordinate system---is
represented by only a small portion of the initial lattice.
Consequently, numerical errors that arise near the outer boundary
dominate the late stages of the evolution.  The second problem is that
the lapse continues to decrease in the vicinity of $r_{\rm EX}$, and
becomes very small.
In this situation truncation errors in quantities near $r_{\rm EX}$
occasionally cause the computation of non-positive values for the
lapse, which immediately leads to code failure.
Both of these problems can be somewhat mitigated by increasing the
numerical resolution.  In harmonic gauge, we were able to simulate the
formation of a black hole and resolve it for about $t\simeq 70
M_{\rm BH}$ using our finest resolution, $N_r=8193$.  However, given these
difficulties induced by the late-time dynamics when using harmonic
coordinates, it is quite natural to try to use the coordinate freedom
provided by the various gauge drivers discussed above to a) attempt to
minimize the time development of the lapse following the formation of
an apparent horizon, and/or b) implement a non-trivial shift vector
with an aim to minimize the outward expansion of $r_{\rm AH}$ at late
times when there is very little matter falling into the black hole.
We thus now summarize our experimentation with several driver
conditions that was focused on realizing these ideas.

As we have already mentioned, one of the main motivations for
Pretorius' development of the driver condition~(\ref{FP_gauge_SS}) was
to keep the lapse from collapsing in the vicinity of
horizons~\cite{FP2}.  Following that work then, we first
used~(\ref{FP_gauge_SS}) to fix the time slicing, while maintaining
harmonic spatial coordinates ($H_r=0$).
However, in contrast to the results reported in~\cite{FP2} (which we
note were performed in three spatial dimensions using  Cartesian
coordinates), we found the evolution in this case to be
significantly less stable than purely harmonic evolution.  For
example, even a small value of $\xi_1$ of order $0.01 M_{\rm BH}^2$
resulted in a code crash at a time about a factor of two earlier than
for the harmonic case, irrespective of the value of the friction
parameter, $\xi_2$.

We next used harmonic slicing, $H_t=0$, while evolving $H_r$ using the
driver~(\ref{FP_gauge_SS_m}).  Here, we found a modest amount of
improvement over the purely harmonic case, in that the
``grid-sucking'' phenomenon described above was slowed, with an
accompanying reduction in the development of numerical error in the
outer, low-resolution region.  For example, the duration of the
evolution of $\Phi_0=3$ initial data with $\xi_{20},\xi_{30} \sim
\O(10) M_{\rm BH}\times (1+5 t^2)^{-1}$, increases by approximately
$20\%$ compared to the corresponding harmonic evolution.

Interestingly, we obtained even better results using certain versions
of the Lindblom {\em et al} gauge drivers.  For the strong-field,
supercritical calculations described here, we found that versions of
the drivers that use the simple scalar operator (\ref{Lind_wave2})
performed better than those that used (\ref{Lind_wave1}).  Moreover,
we found that drivers based on the Bona-Masso slicing and
$\Gamma$-driver shift conditions (with suitably tuned parameters) gave
the best results, and for convenience will hereafter refer to this
specific choice as BMGD.  In particular, relative to other driver
choices, this combination minimized---but unfortunately did not 
completely eliminate---the outward drift of $r_{\rm AH}$.  
Our best configuration allowed
for accurate simulation of black hole spacetimes for about $100 M_{\rm
  BH}$ following the formation of an apparent horizon.  After that
time, code accuracy typically degraded, numerical errors near the
excision became dominant, and a late-time instability ensued.
Based on our experiments, 
it remains unclear whether specific parameter choices for the drivers exist 
that would totally eliminate the drift of the coordinate position 
of the apparent
horizon and, even more importantly, the disastrous collapse of the
lapse inside the horizon.

%%%%%%%%%%%%%%%%%%%%%%%%%%%%%%
\begin{figure}[t!]
  \centering \noindent
  \includegraphics[width=9cm]{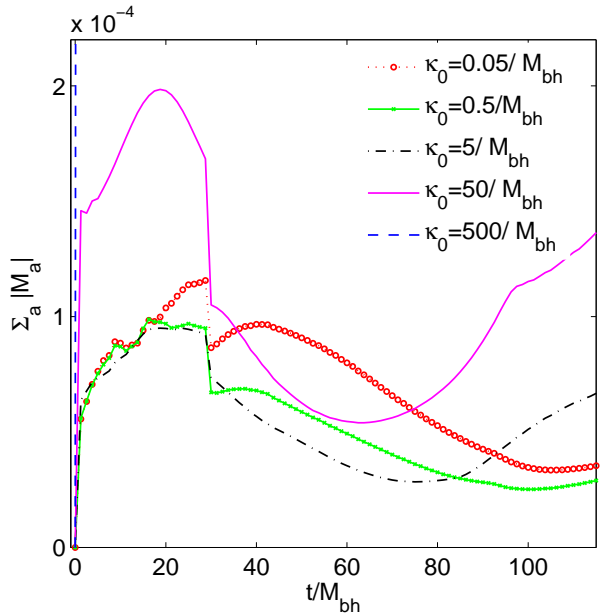}
  \caption[]{ Illustration that the level of constraint preservation
    is dependent on the choice of 
    $\kappa_0$, and that the optimal value is in the range $\kappa_0 \simeq
    M_{\rm BH}^{-1}$. Excessive damping leads to rapid and violent
    growth of the constraints and divergence of the solution. }
  \label{fig_kappa}
\end{figure}
%%%%%%%%%%%%%%%%%%%%%%%%%%%%%%
We now proceed to some details concerning our experience with the BMGD
version of the Lindblom {\em et al} coordinate conditions.
The parameters $q_n$ and $q_s$ that appear in the 
driver definitions---see equations~(\ref{Fn},\ref{Fi})---control 
the relative weight that
the gauge functions, $G^\al$, have in forming the target sources,
$F^\al$.  We also recall that the $G^\al$ vanish when the specific
gauge to which they correspond is attained.  We found it crucial not
to choose $q_n$ too large: usually values in the range $0.01-0.1$
resulted in the most stable evolutions, and would eventually lead to
the desired behavior, $H^\al \rightarrow F^\al$ and $G\rightarrow 0$.
Our implementation was less sensitive to the value of $q_s$, with
results of comparable accuracy and stability being attained for $g_s$
in the range $0.01-10$.

Having determined good values for $q_n$ and $q_s$, we found through
further experimentation that stability is improved when the parameters
$\mu_1,\mu_2$ and $\eta$ are multiplied by a decay factor $ 2M_{\rm
  ADM}/ R$ in the region external to the horizon. This localizes the
effect of the coordinate drivers to the near-horizon region, while
producing a smooth blend to harmonic coordinates at spatial infinity.
In addition, and in analogy to what we did for the subcritical
calculations in generalized harmonic coordinates described earlier in
this section, we further scale $\mu_1$, $\mu_2$ and $\eta$, as well as
$q_n$ by $1/(1+s t^p)$. Here, $s$ and $p$ are again positive tunable
quantities---we typically used $p=2$ and $s=5$---that result in a
late-time decay of the scaled driver parameters.  We note that the
values quoted below generally refer to ``bare'' values for parameters,
with the additional scaling factors being implied.

Fig.~\ref{fig_Fs-Gs_Phi0=3} shows the time development of the
deviation between the target and actual source functions, $F_\alpha$
and $H_\alpha$, respectively, as well as the gauge functions, $G^\al$,
for a typical BMGD calculation.  The computation was performed with
$\mu_1 = 4 \simeq 1/M_{\rm BH}$, $\mu_2=\eta=\eta_2=10$ $\nu=1,
\sigma=-1/3, q_n=0.1$ and $q_s=1$.
The behavior of the two upper plots in the figure reflect the fact
that the $H^\al$ tend to the target source functions soon after an
apparent horizon forms.  Detailed examination of the data reveals that
the match between the target and actual source functions is good
throughout the entire domain for a certain amount of time following
horizon formation. At late times the level of global agreement
degrades, due to large scale variations in the $F^\al$ induced by the
portion of the scalar field that is scattered to infinity.  Despite
this, we still find that actual sources accurately match the targets
on the scale defined by $\mu_1^{-1}\simeq R_{\rm AH}$ (not shown).
The plots of the $L_2$ norms of the gauge functions, $G^\al$, shown in
the bottom half of the figure, reveal a steady decrease in time,
signaling that the desired gauge is being approached asymptotically.

Our investigations of versions of the drivers using target functions
corresponding to ``static'' gauges, such as maximal slicing and
$\Gamma$-freezing, were unsuccessful in the sense that we were not
able to find parameter settings that resulted in $G^\al \rightarrow 0$
as $t\to\infty$.  Interestingly, however, we found that black holes
could nonetheless be simulated using these conditions, with observed
stability properties similar to those obtained using ``dynamic'' gauge
conditions such as BMGD.  This indicates that, at least for the type
of initial data considered here, the stability of the
drivers (\ref{H_driver1},\ref{Q_al1}) does not strongly depend on the
target gauge.

%%%%%%%%%%%%%%%%%%%%%%%%%%%%%%
Finally we note that the use of an appropriate amount of constraint
damping is important for computations in which black holes form.
Fig.~\ref{fig_kappa} shows the behavior of the sum of the $L_2$-norms
of the constraints, (\ref{l2norm}), in a sample run with $N_r=4097$
and using various values for the damping parameter, $\kappa_0$.  The
plots provide clear evidence that the level of constraint maintenance
(as well as the maximum simulation time) is optimized for
$\kappa_0\simeq M_{\rm BH}^{-1}$.  Values of $\kappa_0$ significantly
larger than the optimal value produce rapid code crashes, while those
that are significantly smaller lead to poorer preservation of the
constraints.

%%%%%%%%%%%%%%%%%%%%%%%%%%%%%%%%%%%%%%%
\subsection{Code accuracy, convergence and constraints}
\label{sec_convergence}
%%%%%%%%%%%%%%%%%%%%%%%%%%%%%%%%%%%%%%%

%%%%%%%%%%%%%%%%%%%%%%%%%%%%%%
\begin{figure}[t!]
  \centering \noindent
  \includegraphics[width=9cm]{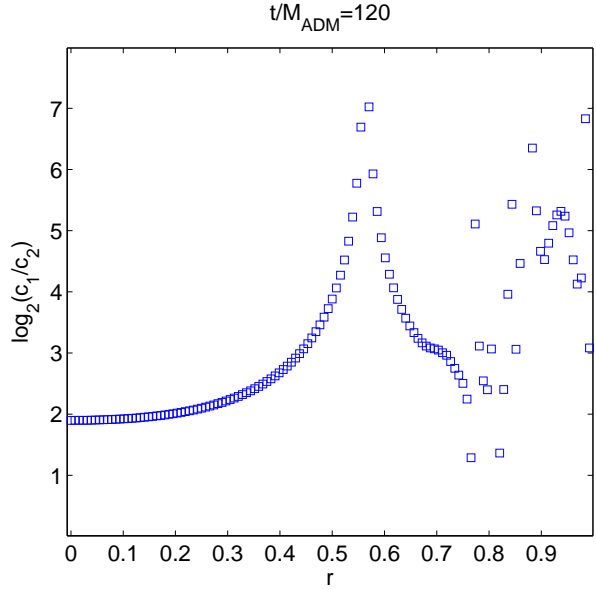}
  \caption[]{Plot showing that the convergence of the scalar field is 
    second order over
    most of the domain, with some irregularities occurring near the
    outer boundary and at the location of the scalar pulse. 
    In
    this simulation, $\Phi_0 =0.55$.}
  \label{fig_logc2c1_weakdata}
\end{figure}
%%%%%%%%%%%%%%%%%%%%%%%%%%%%%%
\begin{figure}[h!]
  \centering \noindent
  \includegraphics[width=11cm]{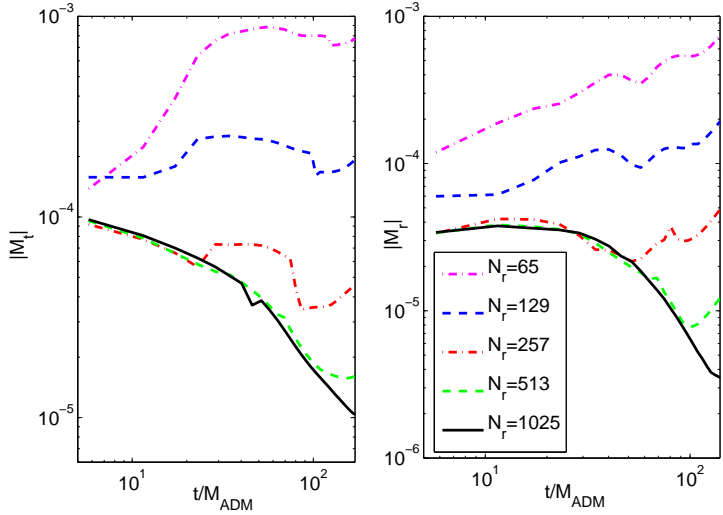}
  \caption[]{A $\log$--$\log$ plot of the $L_2$-norms of the Hamiltonian
    constraint, $M_t$, and the momentum constraint, $M_r$ for 5
    different grid resolutions, and with $\Phi_0 =0.55$. 
    The constraints remain small during
    most of the time-evolution, except for the last moments of the
    simulation, when instabilities set in and eventually lead to code
    failure.  For the most part, the constraints converge as the 
    resolution is increased, but there is a slowing of convergence at 
    the highest resolutions.  This issue is still unresolved, but may 
    be related to the nature of the time iteration.
    }
  \label{fig_logMa_weakdata}
\end{figure}
\begin{figure}[h!]
  \centering \noindent
  \includegraphics[width=8cm]{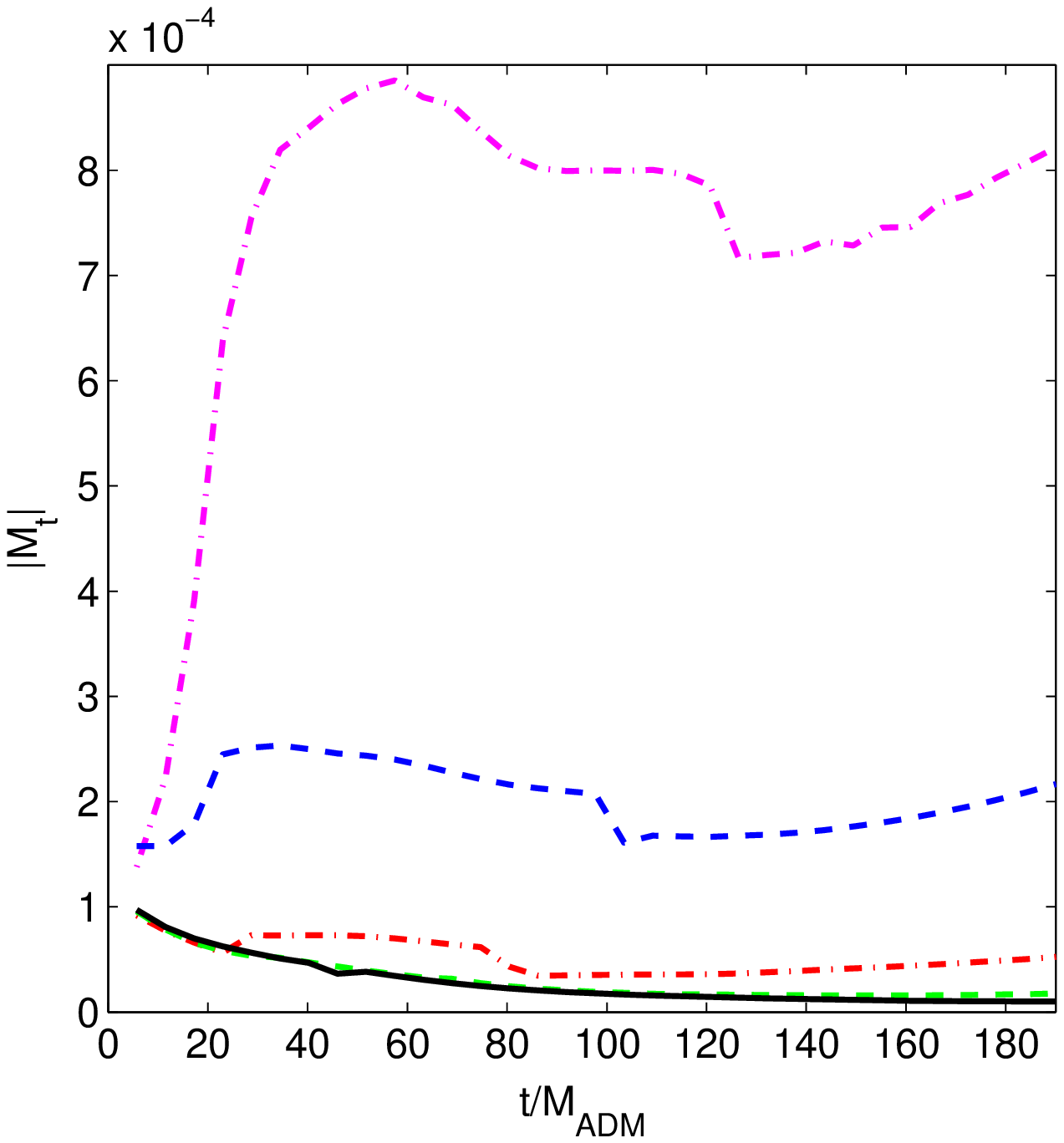}
  \caption[]{Linear plot of the $L_2$-norm of the Hamiltonian
    constraint for 5 different resolutions (the same as those used 
    in Fig.~\ref{fig_logMa_weakdata}), and again with $\Phi_0=0.55$.
    Modulo the comment in the caption of Fig.~\ref{fig_logMa_weakdata},
    second order convergence is observed.}
  \label{fig_Ma_weakdata}
\end{figure}
%%%%%%%%%%%%%%%%%%%%%%%%%%%%%%
In this section we briefly discuss some of the technical issues
relating to the basic performance of our numerical code, including
resolution requirements and checks of convergence.

Not surprisingly, we find that the minimum discretization scale
required to produce an acceptable evolution (for fixed choice of
coordinate conditions) depends on the strength of the initial data.
For example, in the case of weak and intermediate initial data, as
defined previously, even a modest lattice size of $N_r=65$ is enough
to allow for long-time evolution.  However, for stronger data, meshes
sizes of at least $N_r=257$ are required.  Additionally,
our code cannot evolve strong-field data for
arbitrary amounts of coordinate time: generically, numerical problems
develop that lead to a code crash on the order of 10-100 $M_{\rm
  ADM}$, and the precise lifetime of the simulation is dependent
on the strength of the initial data, the resolution, and the details
of the coordinate conditions.

Much of the build-up of error that eventually leads to code failure,
especially in subcritical simulations, can be traced to the use of
spatial compactification.  In all of our calculations, there is
outflux of scalar field to spatial infinity, and as the scalar
radiation propagates to large distances it becomes more poorly
resolved on the mesh, which has uniform spacing in the compactified
radial coordinate.  Untreated, this will lead to spurious reflection
of the waves which will corrupt the interior solution, so we add
Kreiss-Oliger dissipation to explicitly damp the radiation when its
wavelength becomes of order the mesh scale.  Although this damping is
imperfect, we find that increasing the resolution is effective in
extending the lifetime of our evolutions.  As a specific example, for
a calculation which forms a black hole of size ${\tilde R}_{\rm BH}
\simeq 0.6$, and that uses BMGD coordinate conditions and excision, a
grid with $N_r=4097$ is sufficient to keep the reflections small
during all stages of the evolution until $t \simeq 100 M_{\rm BH}$.
Thereafter, an instability appears near $r_{\rm EX}$ and leads to a
code crash.

A crucial test of any finite difference code for the solution of a
system of partial differential equations involves the investigation of
the convergence of the generated numerical solutions as a function of
resolution.  We perform straightforward convergence tests based on the
assumption (originally due to Richardson~\cite{Richardson})
that for any of the unknown functions, $Y(t,r)$, appearing in our
differential system, the corresponding finite difference quantity,
$Y_h(t,r)$
in the limit $h\to0$ admits an asymptotic expansion of the form \be
\label{richardson}
Y_h(t,r) = Y(t,r) + h^p e_p(t,r) + \cdots \ee where $h$ is the
discretization scale, $e_p(t,r)$ is an $h$-independent error function
with smoothness comparable to $Y(t,r)$, and $p$ is an integer which
defines the order of convergence of the scheme.  Following standard
practice, we consider sequences of three calculations performed with
identical initial conditions, but with varying resolutions, $h$, $h/2$
and $h/4$.  We then form the differences, $c_1=Y_{h}-Y_{h/2}$ and
$c_2=Y_{h/2}-Y_{h/4}$, and compute
\be
\label{logc2c1}
\log_2\(\frac{c_1}{c_2}\) \approx p.  \ee
Fig.~\ref{fig_logc2c1_weakdata} shows the results of such a
convergence test for the scalar field, $\Phi(t,r)$, from computations
in pure harmonic coordinates, and with initial data defined by
$\Phi_0\simeq 0.55$.
The plot provides evidence for the
expected second order convergence ($p=2$) of $\Phi_h$, and similar
results are observed for the other dynamical variables.
We note, however, that there is an obvious degradation of convergence 
at the highest resolutions used: this issue has not been resolved, 
but may be related to the time-stepping iteration.

As discussed in Sec.~\ref{sec_numerics}, in the cases where the
Lindblom {\em et al} drivers were used to evolve the source functions,
we used an implicit Euler method to integrate the corresponding finite
difference equations. Since that method is only first-order accurate
in time, the convergence of the overall scheme in only expected to be
first order, and this was in fact observed.

Finally, since we have implemented a free evolution
scheme~\cite{PiranJCP}, we can also assess the convergence of our
numerical solutions by monitoring discrete versions of the Hamiltonian
and momentum constraints, $M_t$ and $M_r$, respectively.  As usual,
these constraints are defined by contracting the Einstein equations
with the unit normal vector to the $t={\rm const.}$ hypersurfaces,
i.e.\ $M_\al \equiv n^\al(G_{\al\bt}-T_{\al\bt})$, where $G_{\al\bt}$
is the Einstein tensor.  In order to estimate how well the constraints
are satisfied, we discretize them to second order, and then compute
their $L_2$-norms, as defined by (\ref{l2norm}), at each time 
step.  Fig.~\ref{fig_logMa_weakdata} shows
a typical plot of the results
for weak initial data ($M_{\rm
  ADM} \simeq 0.01$) evolved with harmonic coordinates.  It is clear
from the figure that the constraint violations remain quite small
during the evolution, and that---modulo the previous remark concerning 
an apparent problem at higher resolutions---the constraints are 
increasingly well
satisfied as $h\to0$.

%%%%%%%%%%%%%%%%%%%%%%%%%%%%%%%%%%%%%%%%%%%%%%%%%
\section{Conclusions}
\label{sec_conclusions}
%%%%%%%%%%%%%%%%%%%%%%%%%%%%%%%%%%%%%%%%%%%%%%%%
We have presented a generalized harmonic formulation of the Einstein
equations for spherically symmetric $D$-dimensional spacetimes. Since it is
natural to choose coordinates in which the symmetries of the geometry are
explicit, we have adopted the usual spherical coordinates.  This 
results in a coordinate singularity at the origin, $r=0$. While
at the continuum level the equations of motion maintain regularity of 
a solution which is initially smooth at the origin, extra care must be 
exercised so that this property is reflected in discrete 
numerical calculations.  We have thus described a procedure to ensure 
that the origin remains regular in numerical calculations, while
preserving the hyperbolicity of the evolution system.

We have investigated the resulting GH system in the context of fully non-linear 
gravitational collapse.  To this end we introduced a real, massless
scalar field, and have used the specification of the initial scalar 
field profile to control the ensuing strength of the gravitational interaction.
The dynamics that we have 
considered range from the dispersion of weak pulses to the collapse of
strong pulses that lead to black hole formation. A key aspect of our 
numerical approach was the use of radial
compactification which, in conjunction with sufficient 
dissipation, provided a viable alternative to the truncation of
the spatial domain and the use of approximate outer boundary 
conditions. Another ingredient of our methodology that was vital
for long-term stability of the numerical calculations was the addition
of constraint-damping terms to the evolution equations.

Our studies of evolutions using several 
coordinate drivers lead us to conclude that, in spherical geometries, the 
gauge drivers discussed in~\cite{FP2,FP3,Lindblometal2,Scheel_etal} are 
less effective relative to the $3+1$ simulations that use 
Cartesian coordinates, and it would be very interesting to understand 
this issue in more detail.
Nevertheless, we found that with a certain amount of parameter tuning
many interesting situations could be successfully simulated with 
drivers that have been proposed in the literature.  Perhaps not 
surprisingly, depending on the situation certain drivers performed better
than others, leading to longer and/or more accurate simulations.
Specifically, the dynamics of weakly gravitating dispersing pulses could
be simulated using any of the considered coordinate choices; however
the pure harmonic gauge arguably provided the cleanest and the simplest
choice. For strong-field data, variations in the performance of the 
various drivers were more apparent.
In particular, for strong but subcritical pulses,
the harmonic gauge quickly lead to coordinate pathologies, signaled by
a collapsing lapse, but this behavior could be partially ameliorated 
by using one 
the drivers given by (\ref{FP_gauge}) and (\ref{FP_gauge_m}). The driver
(\ref{H_driver1}) could also be used to evolve strong-field data in 
some regimes, but the target coordinates which it is designed to
asymptotically enforce, were not achieved, at least not for the range
of the parameters that we explored in this work.

For the case of strong-field, supercritical calculations (i.e.~those 
for which black holes form), we 
found that pure harmonic coordinates could
still be of use.  In the simulations that used excision, it was possible
to evolve black holes for as long as a few tens of dynamical times.
However, the coordinate system remained fairly dynamic even at late
times, leading to collapse of the lapse near the excision surface on
one hand, and to the outwards expansion of the coordinate position of 
the horizon on the other. We were able to use driver conditions to 
moderate the time-dependence, with the best results being obtained 
through the use
of the drivers (\ref{H_driver1}) with the Bona-Masso target slicing
and the $\Gamma$-driven target shift.  It would be very interesting to
find out whether or not parameters and target gauges exist that not
only slow down the time-dependence of the coordinates at late times,
but completely eliminate it.

One of the main goals of this work was to achieve 
a better understanding of the generalized harmonic approach as 
applied to highly symmetric spacetimes, and to prepare ground for 
an exploration of various gravitational phenomena in axisymmetry
using an analogous formalism.  We expect that the
insights gained from our experiments in spherical symmetry will
also prove useful in the axially symmetric case. In particular, coordinates 
that are adapted to the axial symmetry are again formally singular on
the axis, and the equations of motion will need to be regularized there. 
However, the same
regularization described above for spherical symmetry can be readily
extended to that case.  This allows for a regular hyperbolic
formulation in axial symmetry, which will be discussed in 
a subsequent publication~\cite{2Dcode}.

\acknowledgments We would like to thank Frans Pretorius for
interesting and useful discussions, and for tips on the use of the PAMR/AMRD
software \cite{pamr_amrd}.  MWC also gratefully acknowledges the financial 
research support of NSERC and CIFAR, and thanks the MPI-AEI for hospitality
and support while part of this work was carried out.

%%%%%%%%%%%%%%%%%%%%%%%%%%%%%%%%%%%%%%%%%%%%%%%%
\appendix
%%%%%%%%%%%%%%%%%%%%%%%%%%%%%%%%%%%%%%%%%%%%%%%%%
\section{Asymptotically AdS spacetime}
\label{sec_asympt_AdS}
%%%%%%%%%%%%%%%%%%%%%%%%%%%%%%%%%%%%%%%%%%%%%%%%
Here we analyze the asymptotics of $AdS$ spacetime, and discuss a convenient
metric ansatz as well as a normalization of the source functions.

The $AdS_D$ background can be written in the form,
\be
\label{ads1}
ds^2=-(1+\rho^2/\ell^2)d\tau^2 + d\rho^2/(1+\rho^2/\ell^2) +\rho^2
\,d\Omega_n^2, \ee
where $\ell$ is the AdS curvature scale. In our model (\ref{action})
we reproduce asymptotically AdS spacetime by letting $V(0)=\Lambda<0$
that defines $\ell^2 =-(D-1)(D-2)/\Lambda $.  

One of the properties of
the AdS space is that its asymptotic boundary is time-like: in fact,
it takes only a finite time for a light signal to
propagate to the boundary.  Hence, in numerical
implementations, correct treatment of boundary conditions at spatial
infinity is crucial.  To this end it is useful to
transform to conformal coordinates,
\be
\label{ads_conf_coords}
\rho=\ell \tan (r/\ell), ~~~~ \tau=t, \ee
in which the AdS metric becomes
\be
\label{ads2}
ds^2=-\cos^{-2} \( r / \ell\)\(dt^2-dr^2\) + \ell^2\, \tan^2(r/\ell)\,
d\Omega^2_n. \ee
We note that 
the entire space has finite extent $ r \in \[0,\pi \ell /2\]$ in these
coordinates, but that the metric is singular at spatial infinity,
$r=\pi \ell /2$.  

A convenient metric ansatz for evolution using the generalized 
harmonic approach explicitly factors out the background and is given by
\be
\label{ads_ds}
ds^2=-cos^{-2} \( r / \ell\) g_{tt}\, \(dt^2-dr^2\) + 2 g_{tr} dt dr +
\ell^2\, \tan^2(r/\ell)\, e^{2\,S} d\Omega^2_n. \ee
In this case the asymptotic behavior of the fields $g_{ab}$  is regular,
$g_{ab} \rightarrow \eta_{ab}$ and $ S\rightarrow 0 $.

The source function obtained from (\ref{GH_coords}) does not vanish in
spherical coordinates even in pure AdS where it becomes 
\be
H^{AdS}_\mu=\(0, (n/\rho)
[1+((n+2)/n)(\rho^2/\ell^2)]/[1+\rho^2/\ell^2], (n-1)\,\cot
\theta_1,\dots,\cot \theta_{n-1},0\), 
\ee
and where $\rho$ is given in
(\ref{ads_conf_coords}). In analogy with the asymptotically flat case, 
we subtract a
background contribution, 
which is singular at $\rho=0$,
by writing
$H_\al =H_\al + H_\al^{AdS}$, and then use the regular source functions
\be
\label{H_ads}
H_\al=
\(H_t(t,r),H_r(t,r)+\frac{n}{\rho}\frac{1+\frac{n+2}{n}\frac{\rho^2}{\ell^2}}{1+\frac{\rho^2}{\ell^2}}
,(n-1)\cot \theta_1,(n-2)\,\cot\theta_2,\dots,\cot \theta_{n-1},0\).
\ee
%

%%%%%%%%%%%%%%%%%%%%%%%%%%%%%%%%%%%%%%%%%%%%%%%%%
\section{Explicit form of the equations}
\label{sec_SS_eqs}
%%%%%%%%%%%%%%%%%%%%%%%%%%%%%%%%%%%%%%%%%%%%%%%%
We define $g_2\equiv\gtt\, \grr - \gtr^2$, to be the determinant of the
2-metric $g_{ab}$, in (\ref{flat_ds}).  The complex scalar field is
decomposed as $\Phi=\phi_r +i\, \phi_i$.

The Christoffel symbols and the trace of the extrinsic curvature are
given by
\bea
\label{Gammas}
\Gamma_t&=&\(\gtt\, \gtr' -\gtr \, \gtt' -\half \gtt \dot g_{tr}
+\half \grr \dot g_{tt} -n\, g_2\, \dot S \)/g_2\non
\Gamma_r&=&-\frac{n}{r} +\( \half \gtt \,\grr' -\half\grr\,\gtt' -\gtr
\dot g_{rr} + \grr\,\dot g_{tr} - n\,g_2 \, S'\)/g_2  \\
\label{trK}
K&=&\al \(-\frac{n\, \gtr}{r} + \frac{\gtr}{n \grr} \grr' - \gtr' - 2
\gtr \, S' + \half\, \dot g_{rr} +n\, \grr \,\dot S \) /g_2, \eea
The generalized harmonic equations (\ref{EqHgab},\ref{EqHS}) in $4D$ become
\bea
\label{Rtt_4D}
R_{tt}-C_{(t;t)} -\bar T_{tt} &&= \non &&-\frac{1}{4} \(\dot \grr\)^2
(g^{rr})^2+\gtr' \dot \grr (g^{rr})^2+{g^{tt}} \(\dot \gtr\)^2
{g^{rr}}-\frac{1}{2} \gtt'' {g^{rr}}+{g^{tr}} \gtt' \dot \grr
{g^{rr}}+\non && 4 {g^{tr}} \gtr' \dot \gtr
{g^{rr}}+\frac{\(\gtt'\)^2}{2 {g_2}}+\frac{3}{4} (g^{tt})^2 \(\dot
\gtt\)^2-\(\dot \phi_i\)^2-\(\dot \phi_r\)^2-2 \(\dot S\)^2+
\non &&\(\frac{{\gtr} {H_t}-{\gtt} {H_r}}{2
  {g_2}}-\frac{{\gtt}}{{g_2} r}\) \gtt'+\(\frac{2 {\gtt}}{{g_2}
  r}+\frac{{\gtt} {H_r}-{\gtr} {H_t}}{{g_2}}\) \dot \gtr+
\non &&\(\frac{{\grr} {H_t}-{\gtr} {H_r}}{2
  {g_2}}-\frac{{\gtr}}{{g_2} r}\) \dot \gtt+(g^{tr})^2 \gtr' \dot
\gtt+2 {g^{tr}} {g^{tt}} \gtt' \dot \gtt+\frac{1}{2} (g^{tr})^2 \dot
\grr \dot \gtt+\non &&2 {g^{tr}} {g^{tt}} \dot \gtr \dot
\gtt-\dot{H}_t-{g^{tr}} \dot \gtt'-\frac{1}{2} {g^{tt}} \ddot \gtt-{\gtt}
V+2 (g^{tr})^2
\gtt' \dot \gtr,
\eea
%%%%%%%%%%%%%%%%%%%%%%%%%%%%%%%%%%%%%%%%%%%%%%%%%%%%%%%%%%%%%%%%%%%%%%
\bea
\label{Rtr_4D}
R_{tr}-C_{(t;r)}-\bar T_{tr} &&= \non && \frac{1}{2} \grr' \gtr'
(g^{rr})^2+\frac{1}{2} \grr' \dot \grr (g^{rr})^2+{g^{tr}} \(\gtr'\)^2
{g^{rr}}+\frac{1}{2} {g^{tr}} \(\dot \grr\)^2 {g^{rr}}+\non
&&\frac{1}{2} {g^{tr}} \grr' \gtt' {g^{rr}}-\frac{1}{2} \gtr''
{g^{rr}}+{g^{tr}} \grr' \dot \gtr {g^{rr}}+\frac{1}{2} {g^{tr}}
{g^{tt}} \(\gtt'\)^2+{g^{tr}} {g^{tt}} \(\dot \gtr\)^2-\non &&{\gtr}
V+\(\frac{{\grr} {H_t}-{\gtr} {H_r}}{2 {g_2}}- \frac{{\gtr}}{{g_2}
  r}\) \gtt'+\(\frac{(g^{tr})^2}{2}+{g^{rr}} {g^{tt}}\) \gtr'
\gtt'-\frac{1}{2} H_t'+\non &&\(\frac{{\gtt}}{{g_2} r}+\frac{{\gtt}
  {H_r}-{\gtr} {H_t}}{2 {g_2}}\) \dot \grr+\frac{1}{4} \((g^{tr})^2-2
{g^{rr}} {g^{tt}}\) \gtt' \dot \grr+2 {g^{tr}}^2 \gtr' \dot \gtr+\non
&&\(\frac{(g^{tr})^2}{2}+{g^{rr}} {g^{tt}}\) \dot \grr \dot
\gtr+\frac{3}{4} (g^{tr})^2 \grr' \dot \gtt+{g^{tr}} {g^{tt}} \gtr'
\dot \gtt+\frac{1}{2} (g^{tt})^2 \gtt' \dot \gtt+ \non &&\frac{1}{2} (g^{tt})^2 \dot \gtr
\dot \gtt-\frac{1}{2} \dot{H}_r-2 \phi_i' \dot \phi_i-2 \phi_r' \dot
\phi_r-4 S' \dot S-\frac{2 \dot S}{r}-{g^{tr}} \dot \gtr'-\frac{1}{2}
{g^{tt}} \ddot \gtr+\non 
&& \frac{1}{2} {g^{tr}}
{g^{tt}} \dot \grr \dot \gtt, \eea
%%%%%%%%%%%%%%%%%%%%%%%%%%%%%%%%%%%%%%%%%%%%%%%%%%%%%%%%%%%%%%%%%%%%%%
\bea
\label{Rrr_4D}
R_{rr}-C_{(r;r)}-\bar T_{rr} &&= \non && \frac{1}{2} \grr' \gtt'
(g^{tr})^2+2 \gtr' \dot \grr (g^{tr})^2+\grr' \dot \gtr (g^{tr})^2+2
{g^{rr}} \grr' \gtr' {g^{tr}}+2 {g^{rr}} \grr' \dot \grr {g^{tr}}+\non
&& {g^{tt}} \gtt' \dot \grr {g^{tr}}+4 {g^{tt}} \gtr' \dot \gtr
{g^{tr}}-\dot \grr' {g^{tr}}+\frac{3}{4} (g^{rr})^2
\(\grr'\)^2+{g^{rr}} {g^{tt}} \(\gtr'\)^2-\non &&\(\phi_i'\)^2-\(\phi_r'\)^2-2
\(S'\)^2+\frac{\(\dot \grr\)^2}{2 {g_2}}-{\grr}
V+\(\frac{{\gtt}}{{g_2} r}+\frac{{\gtt} {H_r}-{\gtr} {H_t}}{2 {g_2}}\)
\grr'+\non &&\(\frac{{\grr} {H_t}-{\gtr} {H_r}}{{g_2}}-\frac{2
  {\gtr}}{{g_2} r}\) \gtr'- H_r'-\frac{4 S'}{r}+\(\frac{{\gtr}}{{g_2}
  r}+\frac{{\gtr} {H_r}-{\grr} {H_t}}{2 {g_2}}\) \dot \grr+\non
&&({g^{tt}})^2 \gtt' \dot \gtr- \frac{1}{4}
(g^{tt})^2\(\gtt'\)^2-\frac{1}{2} {g^{tt}}\ddot
\grr-\frac{1}{2} {g^{rr}} \grr'' , \eea
%%%%%%%%%%%%%%%%%%%%%%%%%%%%%%%%%%%%%%%%%%%%%%%%%%%%%%%%%%%%%%%%%%%%%%
\bea
\label{Rtthth_4D}
R_{\th\th}-C_{(\th;\th)}-\bar T_{\th\th} &&= \non && -\frac{2 S'
  {\gtt}}{{g_2} r}-\frac{{\gtt}}{{g_2} r^2}+ \frac{e^{-2 S}}{r^2}-V+
\frac{2 {\gtr} \dot S}{{g_2} r}+{H_t} \(\frac{S'
  {\gtr}}{{g_2}}+\frac{{\gtr}}{{g_2} r}-\frac{{\grr} \dot
  S}{{g_2}}\)+\non &&{H_r} \(-\frac{S'
  {\gtt}}{{g_2}}-\frac{{\gtt}}{{g_2} r}+\frac{{\gtr} \dot
  S}{{g_2}}\)-2 {g^{tr}} \dot S'-{g^{tt}} \ddot S-{g^{rr}} S''.  \eea
Written in full, 
the constraint damping terms, $Z_{\mu\nu}=\kappa \( n_{(\mu}\cC_{\nu)}
-\half g_{\mu\nu} \, n^\bt \, \cC_\bt \)$, that we subtract from the
above equations to form (\ref{Eqs_constrdamp}), are
\bea
\label{Cdmp_4D}
Z_{tt} &=& \non &&\frac{\alpha \kappa}{g_2} \Big[-\frac{{\gtr} \grr'
  {\gtt}^2}{4 \ {g_2}}+\frac{{\grr} \dot \grr {\gtt}^2}{4
  {g_2}}-\frac{{\gtr} {H_r} {\ \gtt}}{2}-\frac{\({\grr} {\gtt}-2
  {\gtr}^2\) \gtr' {\gtt}}{2 {g_2}}+\non &&{\ \gtr} S'
{\gtt}-\frac{{\grr} {\gtr} \dot \gtr {\gtt}}{2 \
  {g_2}}+\({\gtr}^2-\frac{{\grr} {\gtt}}{2}\) {H_t}+\frac{{\gtr} \(3 \
  {\grr} {\gtt}-4 {\gtr}^2\) \gtt'}{4 {g_2}}-\non &&\frac{{\grr}
  \({\grr} \ {\gtt}-2 {\gtr}^2\) \dot \gtt}{4 {g_2}}+\({\grr} {\gtt}-2
{\gtr}^2\) \ \dot S \Big], \eea \bea
%%%%%%%%%%%%%%%%%%%%%%%%%%%%%%%%%%%%%%%%%%%%%%%%%%%%%%%%%%%%%%%%%%%%%%
Z_{tr} &=& \non &&\frac{\alpha \kappa}{g_2} \Big[-\frac{{\gtt} \dot
  \gtr {\grr}^2}{2 \ {g_2}}+\frac{{\gtr} \dot \gtt {\grr}^2}{4
  {g_2}}-\frac{{\gtt} {H_r} {\ \grr}}{2}+\frac{{\gtr} {H_t}
  {\grr}}{2}-\frac{{\gtt}^2 \grr' \ {\grr}}{4 {g_2}}+\non
&&\frac{{\gtr} {\gtt} \gtr' {\grr}}{2 {g_2}}+\frac{\({\ \grr} {\gtt}-2
  {\gtr}^2\) \gtt' {\grr}}{4 {g_2}}+{\gtt} S' \ {\grr}+\frac{{\gtr}
  {\gtt} \dot \grr {\grr}}{4 {g_2}}-{\gtr} \dot S {\ \grr}\Big], \eea
\bea
%%%%%%%%%%%%%%%%%%%%%%%%%%%%%%%%%%%%%%%%%%%%%%%%%%%%%%%%%%%%%%%%%%%%%%
Z_{rr} &=& \non &&\frac{\alpha \kappa}{g_2} \Big(-\frac{{\gtt} \dot
  \gtr {\grr}^2}{2 \ {g_2}}+\frac{{\gtr} \dot \gtt {\grr}^2}{4
  {g_2}}-\frac{{\gtt} {H_r} {\ \grr}}{2}+\frac{{\gtr} {H_t}
  {\grr}}{2}-\frac{{\gtt}^2 \grr' \ {\grr}}{4 {g_2}}+\non
&&\frac{{\gtr} {\gtt} \gtr' {\grr}}{2 {g_2}}+\frac{\({\ \grr} {\gtt}-2
  {\gtr}^2\) \gtt' {\grr}}{4 {g_2}}+{\gtt} S' \ {\grr}+\frac{{\gtr}
  {\gtt} \dot \grr {\grr}}{4 {g_2}}-{\gtr} \dot S {\ \grr}\Big], \eea
\bea
%%%%%%%%%%%%%%%%%%%%%%%%%%%%%%%%%%%%%%%%%%%%%%%%%%%%%%%%%%%%%%%%%%%%%%
Z_{\th\th} &=& \non &&\frac{\alpha \kappa}{g_2} \Big[\frac{\dot \gtt
  {\grr}^2}{4 \ {g_2}}+\frac{{H_t} {\grr}}{2}+\frac{{\gtt} \gtr'
  {\grr}}{2 \ {g_2}}-\frac{{\gtr} \gtt' {\grr}}{4 {g_2}}-\non
&&\frac{{\gtr} \dot \gtr \ {\grr}}{2 {g_2}}-\dot S {\grr}-\frac{{\gtr}
  {H_r}}{2}-\frac{{\gtr} \ {\gtt} \grr'}{4 {g_2}}+{\gtr} S'+\frac{\(2
  {\gtr}^2-{\grr} {\gtt}\) \ \dot \grr}{4 {g_2}}\Big].  \eea
%
% The scalar-field equation in (\ref{}) splits in to two equations for
% real and imaginary components, and each equation reads
% %
% \bea
% \label{boxphi_4D}

% \eea
% %
Finally, the Hamiltonian and momentum constraints, $M_\al \equiv
n^\al(G_{\al\bt}-T_{\al\bt})$, take the form
\bea
\label{Mal_4D}
M_t&=&-\frac{1}{2} {g_2} {\gtt} \(\phi_i'\)^2-\frac{1}{2} {g_2} {\gtt}
\(\phi_r'\)^2-3 {g_2} {\gtt} \(S'\)^2+\frac{1}{2} {g_2} {\grr} \(\dot
\phi_i\)^2+\frac{1}{2} {g_2} {\grr} \(\dot \phi_r\)^2-\non &&{g_2}
{\grr} \(\dot S\)^2-\frac{{g_2} \({g_2} V r^2-e^{-2 S}
  {g_2}+{\gtt}\)}{r^2}-2 {g_2} {\gtt} S''+\frac{4 \({\grr} {\gtr} {\gtt}-{\gtr}^3\) \dot S}{r}+\non &&\grr'
\({g^{rr}}^2 S' {g_2}^2+{g^{rr}} {g^{tr}} \dot S
{g_2}^2+\frac{{\gtt}^2}{r}\)+ S' \(-\frac{6 {g_2} {\gtt}}{r}-8
\({\gtr}^3+{g_2}^2 {\grr} {g^{rr}} {g^{tr}}\) \dot S\)+\non &&\gtr'
\(2 {g^{rr}} {g^{tr}} S' {g_2}^2+2 {g^{rr}} {g^{tt}} \dot S
{g_2}^2-\frac{2 {\gtr} {\gtt}}{r}\)+\gtt' \({g^{tr}}^2 S'
{g_2}^2+{g^{tr}} {g^{tt}} \dot S {g_2}^2+\frac{{\gtr}^2}{r}\)+\non
&&\(2 {\grr} {\gtr} {\gtt}-2 {\gtr}^3\) \dot S'-{g_2} \dot \grr \dot
S.  \eea \bea
%%%%%%%%%%%%%%%%%%%%%%%%%%%%%%%%%%%%%%%%%%%%%%%%%%%%%%%%%%%%%%%%%%%%%%%%%%
M_r&=&\({\gtr}^3+{g_2}^2 {\grr} {g^{rr}} {g^{tr}}\) \(\phi_i'\)^2+2
{g_2} { \grr} \dot \phi_i \phi_i'+\({\gtr}^3+{g_2}^2 {\grr} {g^{rr}}
{g^{tr}}\) \(\phi_r'\)^2-\non &&2 {g_2} {\gtr}
S''+\(-S' {g_2}- \frac{{g_2}}{r}\) \dot \grr+2 {g_2} {\grr} \phi_r'
\dot \phi_r+\frac{2 {g_2} {\grr} \dot S}{r}+2 {g_2} { \grr} \dot
S'+\non &&S' \(\frac{4 \({\gtr}^3-{\grr} {\gtr} {\gtt}\)}{r}+4 {g_2}
{\grr} \dot S\)+\grr' \(-{g^{rr}} {g^{tr}} S' {g_2}^2-{g^{tr}}^2 \dot
S {g_2}^2+\frac{{\gtr} {\gtt}}{r}\)+\non &&\gtr' \(-2 {g^{tr}}^2 S'
{g_2}^2-2 {g^{tr}} {g^{tt}} \dot S {g_2}^2-\frac{2
  {\gtr}^2}{r}\)-2 {g_2} {\gtr} \(S'\)^2+\non
&&\gtt' \(-{g^{tr}} {g^{tt}} S' {g_2}^2-{g^{tt}}^2 \dot
S {g_2}^2+\frac{{\grr} {\gtr}}{r}\).  \eea
%

%%%%%%%%%%%%%%%%%%%%%%%%%%%%%%%%%%%%%%%%%%%%%%%%
\section{Discretization}
\label{sec_discrt}
%%%%%%%%%%%%%%%%%%%%%%%%%%%%%%%%%%%%%%%%%%%%%%%%
%
The second order accurate finite difference approximations (FDAs) for
the time derivatives on a uniform grid with spacings $\Delta r, \Delta
t$ at a point $(n,i)$ (see Fig.~\ref{fig_grid}) are
\bea
\label{FDA_t}
\pa_t Y_i^n &=& \frac{Y_{i}^{n+1} -Y_{i}^{n-1}}{2\,\Delta t}, \non
\pa^2_t Y_i^n &=& \frac{Y_{i}^{n+1} -2\, Y_i^n +Y_{i}^{n-1}}{(\Delta
  t)^2} . \eea
Here ``second order'' means that the continuum expression is approached
by the FDA counterpart at a rate $ \O(\Delta t^2)$.  For the spatial
and mixed derivatives the stencil is modified depending on the
position of the mesh point relative to the extremities of the grid. 
We use second order accurate expressions of the form
\begin{itemize}
\item Centered derivative.
  \bea
  \label{FDA_c}
  \pa_r Y_i^n &=& \frac{Y_{i+1}^n -Y_{i-1}^n}{2\,\Delta r}, \non
  \pa^2_r Y_i^n &=& \frac{Y_{i+1}^n -2\, Y_i^n +Y_{i-1}^n}{(\Delta
    r)^2}, \non \pa^2_{rt} Y_i^n &=& \frac{Y_{i+1}^{n+1} -
    Y_{i+1}^{n-1} -Y_{i-1}^{n+1}+Y_{i-1}^{n-1}}{4\, \Delta r\, \Delta
    t} \eea

\item One-sided (backward) derivative.
  \bea
  \label{FDA_b}
  \pa_r Y_i^n &=& \frac{4\,Y_{i+1}^n -3\,Y_{i}^n- Y_{i+2}^n}{2\,\Delta
    r}, \non \pa^2_r Y_i^n &=& \frac{2\,Y_{i}^n -5\, Y_{i+1}^n
    +4\,Y_{i+2}^n-Y_{i+3}^n}{(\Delta r)^2}, \non \pa^2_{rt} Y_i^n &=&
  \frac{4\,Y_{i+1}^{n+1} -3\,Y_{i}^{n+1}-
    Y_{i+2}^{n+1}-4\,Y_{i+1}^{n-1}+3\,Y_{i}^{n-1}+ Y_{i+2}^{n-1}}{4\,
    \Delta r\, \Delta t}, \ \eea

\end{itemize}
%%%%%%%%%%%%%%%%%%%%%%%%%%%%%%%%%%%%%%%%%%%%%%%%%%%%%%%%%%%%%%%%%%%%%%%%

\end{document}